\documentclass[
twocolumn,
aps,
prb,
reprint,
superscriptaddress,
amsmath,amssymb
]{revtex4-1}
\usepackage[pdftex]{graphicx} 
\usepackage[hidelinks]{hyperref} 
\hypersetup{
 setpagesize=false,
 bookmarksnumbered=true,%
 bookmarksopen=true,%
 colorlinks=true,%
 linkcolor=blue,
 citecolor=blue,
 urlcolor=blue
}

\usepackage{braket}
\usepackage{times,multirow,amsfonts,bm,xspace,pifont,mathtools}
\usepackage[normalem]{ulem}
\usepackage{soul}
\begin{document}
\newcommand{\rr}{{\bm r}}
\newcommand{\q}{{\bm q}}
\renewcommand{\k}{{\bm k}}
\newcommand*\wien    {\textsc{wien}2k\xspace}
\newcommand*\textred[1]{\textcolor{red}{#1}}
\newcommand*\textblue[1]{\textcolor{blue}{#1}}
\newcommand*\YY[1]{\textcolor{red}{#1}}
\newcommand*\KN[1]{\textcolor{red}{#1}}
\newcommand*\MS[1]{\textcolor{magenta}{#1}}
\newcommand*\JI[1]{\textcolor{red}{#1}}
\newcommand{\JIS}[1]{\textcolor{cyan}{\sout{#1}}}
\newcommand{\ki}[1]{{\color{red}\st{#1}}}
\newcommand{\YYS}[1]{\textcolor{blue}{\sout{#1}}}

\title{Correlation-induced Fermi surface evolution and topological crystalline superconductivity in \texorpdfstring{CeRh$_2$As$_2$}{CeRh2As2}} 

\author{Jun Ishizuka}
\affiliation{Faculty of Engineering, Niigata University, Ikarashi, Niigata 950-2181, Japan}
\affiliation{Institute for Theoretical Physics, ETH Z\"{u}rich, 8093 Z\"{u}rich, Switzerland}

\author{Kosuke Nogaki}
\affiliation{Department of Physics, Graduate School of Science, Kyoto University, Kyoto 606-8502, Japan}

\author{Manfred Sigrist}
\affiliation{Institute for Theoretical Physics, ETH Z\"{u}rich, 8093 Z\"{u}rich, Switzerland}

\author{Youichi Yanase}
\affiliation{Department of Physics, Graduate School of Science, Kyoto University, Kyoto 606-8502, Japan}
\date{\today}

\begin{abstract}
Locally noncentrosymmetric structures in crystals are attracting much attention owing to emergent phenomena associated with the sublattice degree of freedom. The newly discovered heavy fermion superconductor CeRh$_2$As$_2$ is considered to be an excellent realization of this class. Angle-resolved photoemission spectroscopy experiments recently observed low-energy spectra of electron and hole bands and characteristic Van Hove singularities, stimulating us to explore the electronic correlation effect on the band structure. In this Letter, we theoretically study the electronic
state and topological superconductivity from first principles. Owing to the Coulomb repulsion $U$ of Ce 4$f$ electrons, the low-energy band structure is modified in accordance with the experimental result. We show that Fermi surfaces change significantly from a complicated three-dimensional structure to a simple two-dimensional one. Fermi surface formulas for one-dimensional $\mathbb{Z}_2$ invariants in class D indicate topological crystalline superconductivity protected by the glide symmetry in a broad region for $U$.
The classification of superconducting gap structure reveals topologically protected excitation gap and node. 
Our findings of the correlation-induced evolution of electronic structure provide a basis to clarify the unusual phase diagram of CeRh$_2$As$_2$ including superconductivity, magnetic order, and quadrupole density wave, and accelerate the search for topological superconductivity in strongly correlated electron systems. 
\end{abstract}

\maketitle

\section{Introduction}
The locally non-centrosymmetric superconductor CeRh$_2$As$_2$ stands out because it shows two distinct superconducting phases under magnetic fields directed along the $c$ axis \cite{Khim2021Field-induced}.
The upper critical field reaches 14 T at zero temperature, significantly exceeding the Pauli paramagnetic limit in view of the low critical temperature, $T_{\rm c} \approx 0.35$ K.
Rotating the field towards the $ab$ plane quickly suppresses the critical field down to 2 T, and the high-field superconducting phase disappears~\cite{Landaeta2022Field-Angle}.
This high-field phase is consistently described as a pair-density wave (PDW) with an order parameter switching from a uniform even to a uniform odd-parity pairing state\cite{Yoshida2012Pair-density, Fischer2023Superconductivity, Khim2021Field-induced, Landaeta2022Field-Angle, Ogata2023Parity}. 

The unusual superconducting phase diagram seems to emerge from strongly correlated electrons.
The specific heat coefficient is enhanced as $\gamma\sim1000$ mJ/molK$^2$ indicating strong influence by Ce $4f$ electrons, and superconductivity occurs from the resistivity $\rho\propto\sqrt{T}$ suggesting the quantum criticality \cite{Khim2021Field-induced}.
Intriguingly, a nonmagnetic quadrupole density wave above $T_{\rm c}$ has been proposed based on the resistivity, specific heat, and thermal expansion measurements \cite{Khim2021Field-induced, Hafner2022Possible, Mishra2022Anisotropic, Semeniuk2023Decoupling}.
The appearance of antiferromagnetic order within the superconducting phase has also been suggested by nuclear quadrupole and magnetic resonance measurements \cite{Kibune2022CeRh2As2, Kitagawa2022Two-Dimensional, Ogata2023}, in which magnetic moments are thought to break global inversion symmetry in the system.

Theoretical studies of CeRh$_2$As$_2$ have provided further insight into superconductivity and electronic states \cite{Schertenleib2021Unusual, Skurativska2021Spin, Nogaki2021topological, Mockli2021Superconductivity, Ptok2021Electronic, Cavanagh2022Nonsymmorphic, Nogaki2022Even-odd, Hazra2022Triplet, Machida2022Violation, Onishi2022Low-Temperature, Amin2023Kramers}.
The crystal structure has an inversion center between two Ce atoms but not on the Ce atom. We refer to such a property as local inversion symmetry breaking. 
Due to the lack of inversion symmetry on the Ce atom, staggered antisymmetric spin-orbit coupling (ASOC) plays a central role.
The impact of such staggered ASOC depends on the ASOC and the interlayer hopping ratio and can be enhanced if the latter is weak~\cite{Yoshida2012Pair-density,sumita2016}. Indeed, Rh$_2$As$_2$ block layers considerably reduce interlayer hopping between the Ce atoms, and the nonsymmorphic crystal symmetry exactly prohibits this hopping at the Brillouin zone boundary \cite{Cavanagh2022Nonsymmorphic,sumita2016}.
As a consequence, unusual $H$-$T$ phase diagrams involving even and odd-parity superconducting phases are highly expected \cite{Skurativska2021Spin, Mockli2021Superconductivity, Cavanagh2022Nonsymmorphic, Nogaki2022Even-odd}.

The low-energy electronic structure near the Fermi level is crucial to verify these scenarios. 
First-principles calculations show that the low-energy states are well described by the Ce $4f$ bands hybridizing with Rh $4d$ bands \cite{Nogaki2021topological, Ptok2021Electronic, Hafner2022Possible, Cavanagh2022Nonsymmorphic}, consistent with the heavy fermion behavior observed in experiments.
However, the optical conductivity measured in experiment quantitatively deviates from the simple hybridization picture and suggests the importance of the correlation effects \cite{Kimura2021Optical}.
Renormalized band structure calculations have been conducted to gain insight into the electronic state \cite{Hafner2022Possible, Cavanagh2022Nonsymmorphic}.
 The hole and electron Fermi surfaces (FSs) are elliptical and cylindrical in the renormalized band calculation, whereas they have a complex three-dimensional (3D) structure in the standard bare band calculation.
It should be noted that the magnetic field separating the low- and high-field superconducting phases $H^*$ reaches $H^*/T_{\rm c} \sim 10$ in CeRh$_2$As$_2$~\cite{Khim2021Field-induced}, which is consistent rather with the 
fluctuation exchange analysis $\sim15$~\cite{Nogaki2022Even-odd} than with the mean-field value $\sim2$~\cite{Yoshida2012Pair-density}. This fact also implies the presence of strong correlation effects in CeRh$_2$As$_2$. 

Studies of CeRh$_2$As$_2$ may impact research on topological superconductivity because designing topological materials is a central issue in modern condensed matter physics.
Thanks to the recent topological classification theory, several setups for topological superconductivity have been identified \cite{Qi2011,Sato2016,Sato2017,Schnyder2008,Kitaev2009,Ryu2010}.
The concept of topological crystalline superconductivity (TCSC), in which the nontrivial topology is protected by crystalline symmetry, has opened up a vast area of research \cite{Fu2011,Zhang2013,Chiu2013,Morimoto2013,Shiozaki2014,Chiu2014,Ueno2013,Tsutsumi2013,Yoshida2015,Fang2015,Shiozaki2015,Shiozaki2016,Shapourian2018,Yanase2017,Daido2019,Ono2019,Ono2020,Ono2020_2,Skurativska2020,Geier2020,Shiozaki2019,Ahn2020}.  For example, it has been proposed that UPt$_3$ \cite{Yanase2017}, UCoGe under pressure \cite{Daido2019}, and the PDW state in CeRh$_2$As$_2$ \cite{Nogaki2021topological} can be topological crystalline superconductors protected by the nonsymmorphic glide symmetry.
For CeRh$_2$As$_2$, topological $\mathbb{Z}_2$ invariants defined in the glide invariant ${\bm k}$ space can be calculated by counting the number of FSs on the high symmetry $\Gamma$-$M$ and $Z$-$A$ lines. 
Indeed, the previous study concluded the TCSC from a first-principles calculation~\cite{Nogaki2021topological}. However, it is widely known that the first-principles band calculations may be  insufficient to predict FSs of $f$-electron systems because of the strong correlation, e.g., CeCoIn$_5$ \cite{Haule2010Dynamical,Nomoto2014Fermi}, CeIrIn$_5$ \cite{Choi2012Temperature}, CeCu$_2$Si$_2$ \cite{Ikeda2015Emergent}, and UTe$_2$ \cite{Ishizuka2019Insulator-Metal,Xu2019Quasi-Two-Dimensional}. Therefore, clarifying the topological nature of CeRh$_2$As$_2$ with the strong correlation among the $f$ electrons is highly desired.

In this work, we investigate the electronic state in paramagnetic CeRh$_2$As$_2$ with the electronic correlation effects by the density functional theory plus Hubbard $U$ (DFT$+U$) calculations. 
We show the dimensional crossover of the FSs from 3D to two dimension (2D) due to the correlation effects.
The band dispersion for intermediate $U$ is in good agreement with the recent angle-resolved photoemission spectroscopy (ARPES) experiment \cite{Chen2023Coexistence, Wu2023Quasi}.
Linking the FS topology to the FS formula of topological invariants, we show that the TCSC is stable in a wide region for the interaction parameter $U$.
In addition, the superconducting gap structures are predicted in terms of symmetry and topology.
Combining the results, the low-energy excitation in the bulk and the Majorana states on the surface are predicted for all the odd-parity superconducting states.

\section{Result}
\subsection{Band calculations}
The atom-resolved band structures by the DFT$+U$ method are summarized in Fig.~\ref{fig:band_u0-5}.
Details of our band calculation are given in the Supplemental Material \cite{suppl}.
The result for $U=0$ eV is in good agreement with the previous reports \cite{Nogaki2021topological, Ptok2021Electronic}. 
As expected with the usual DFT$+U$ calculation, introducing $U$ lifts some empty Ce $4f$ bands. 
The low-energy band dispersion on the $\Gamma$-$X$-$M$-$\Gamma$ line is well fitted by the ARPES data~\cite{Chen2023Coexistence} as shown later. 
We observe two remarkable features. 
First, a Dirac point at the $X$ point near $-0.3$ eV is shifted up toward the Fermi level ($\sim-0.13$ eV for $U=4$ eV), qualitatively consistent with the recent ARPES measurement, although the ARPES observed it slightly closer to the Fermi level $\sim-75\pm60$ meV. The electron bands are  fourfold degenerate at the $X$ point because the nonsymmorphic space group symmetry $P4/nmm$ of CeRh$_2$As$_2$ protects the degeneracy and stabilizes the Dirac electrons~\cite{Chen2023Coexistence}. 
As the Dirac point approaches the Fermi level, a hybridization between Rh and Ce atoms becomes noticeable, resulting in a Van Hove singularity hybridizing with $4f$ electrons close to the Fermi level, consistent with Ref.~\onlinecite{Chen2023Coexistence}. Second, electron pockets constructed by the $4f$ electron bands appear around the $M$ and $A$ points for $U\geq1$ eV, involving a change of orbital characters from Rh(1) to Ce.
Owing to the flat band properties, both the Van Hove singularity and heavy electron bands may play a significant role for magnetic order, quadrupole order, and superconductivity.

\begin{figure*}[tbp]
\includegraphics[width=1.0\linewidth]{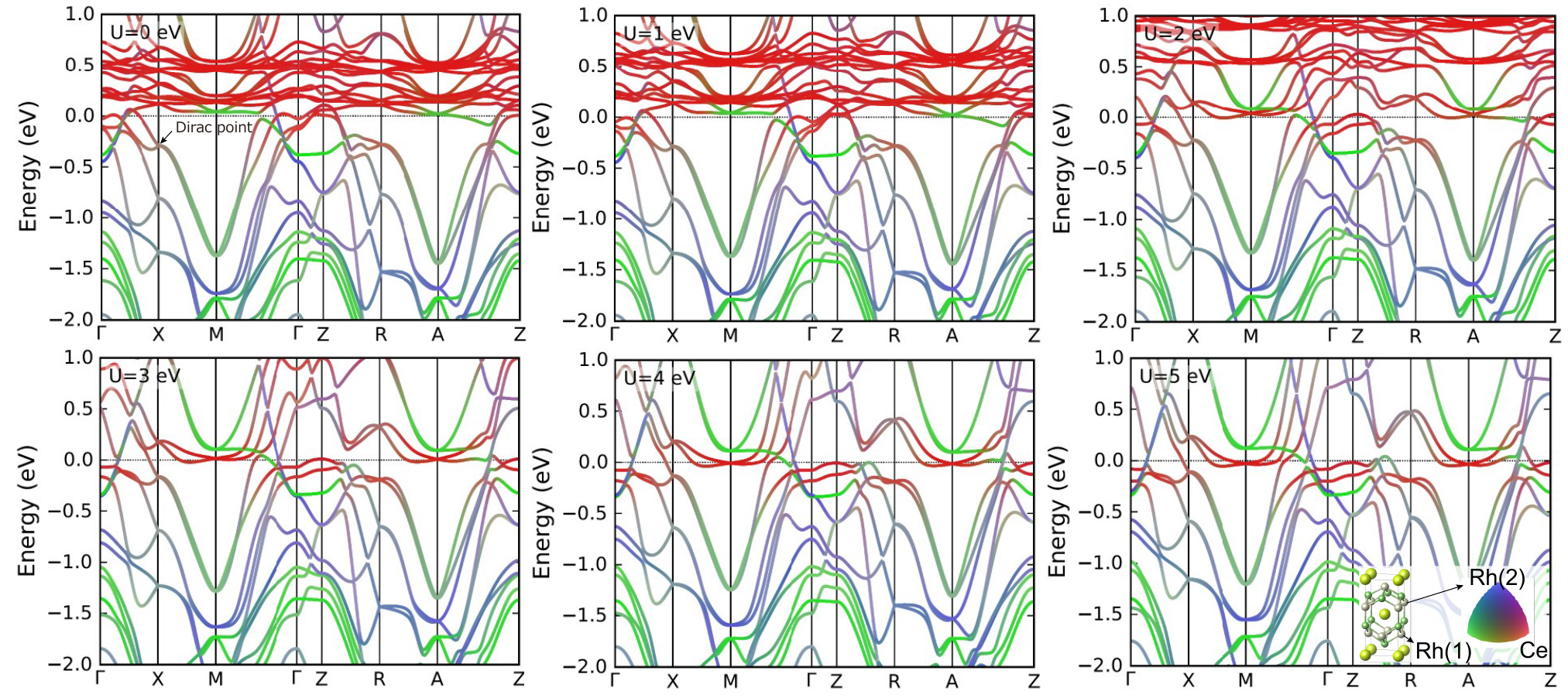}
\centering
\caption{Band structure along the high symmetry lines by the DFT$+U$ calculation with various values of $U$. The red, green, and blue colors represent the weight of atoms for Ce, Rh(1) and Rh(2), respectively.
\label{fig:band_u0-5}}
\end{figure*}

\begin{figure}[tbp]
\includegraphics[width=1.0\linewidth]{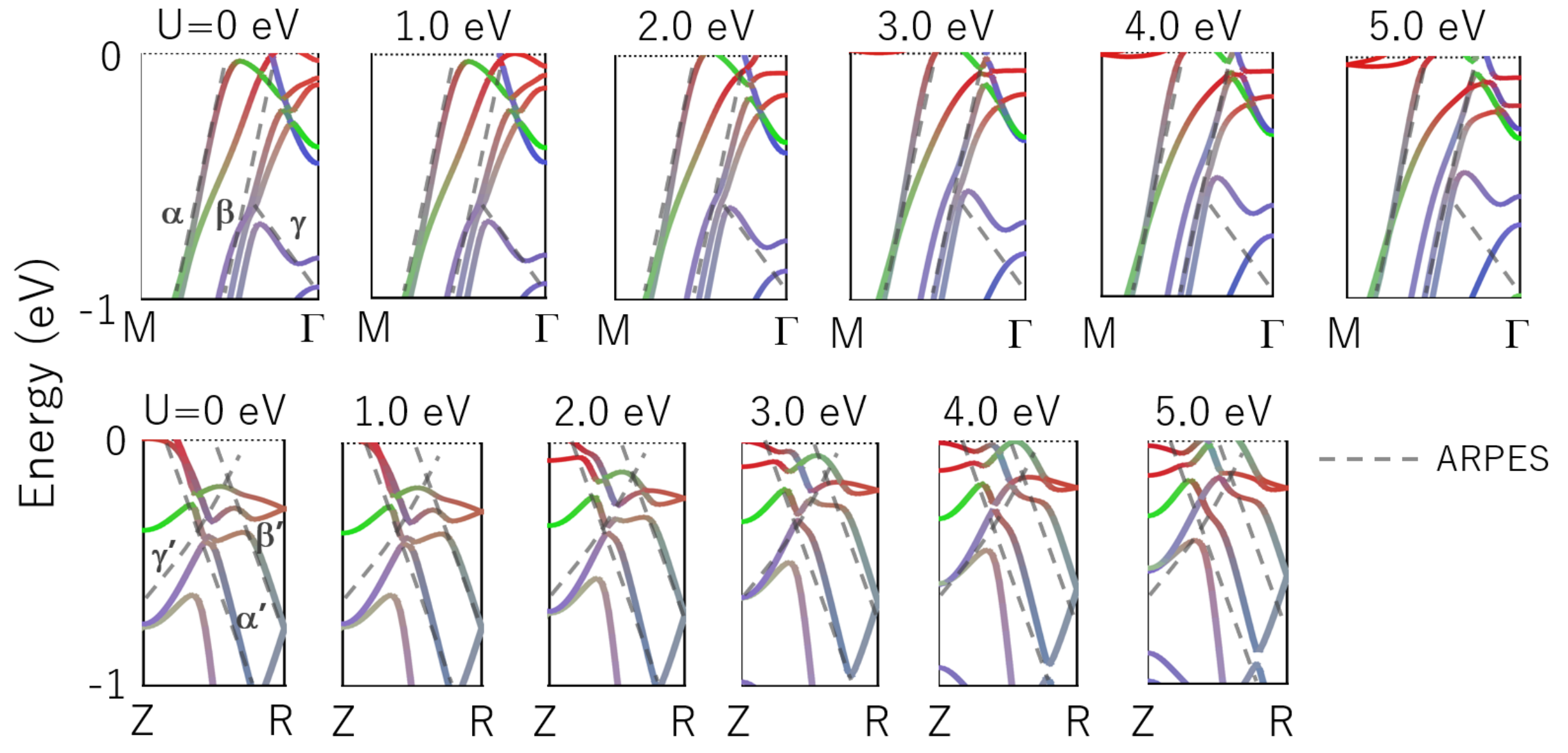}
\centering
\caption{Band structure by the DFT$+U$ calculation along the $M$-$\Gamma$ and $Z$-$R$ lines. Dashed lines are extracted from the ARPES experiment \cite{Chen2023Coexistence}.
\label{fig:band_ARPES}}
\end{figure}

Figure \ref{fig:band_ARPES} compares the DFT$+U$ band structure with the ARPES data along the $M$-$\Gamma$ and $Z$-$R$ lines. ARPES has observed $\alpha(\alpha')$, $\beta(\beta')$, and $\gamma(\gamma')$ branches. The DFT$+U$ results actually capture these branches. Importantly, the $4f$ orbital character is dominant near the Fermi level. Thus, the agreement between the calculation and experiment is found in not only Rh(1) and Rh(2) components but also Ce components, implying the accuracy of the DFT$+U$ approximation. However, we can see some deviations in $\beta'$ and $\gamma'$, which do not cross the Fermi level for $U\leq3$ eV.  Further experimental and theoretical studies are desired for comparison, e.g., considering the self-energy correction from the localized $f$-electrons by DFT$+$DMFT. This is beyond the scope of this paper, and we left it for further study. For experiments, accurate observation of the bands close to the Fermi level is desirable.

\begin{figure}[tbp]
\includegraphics[width=1.0\linewidth]{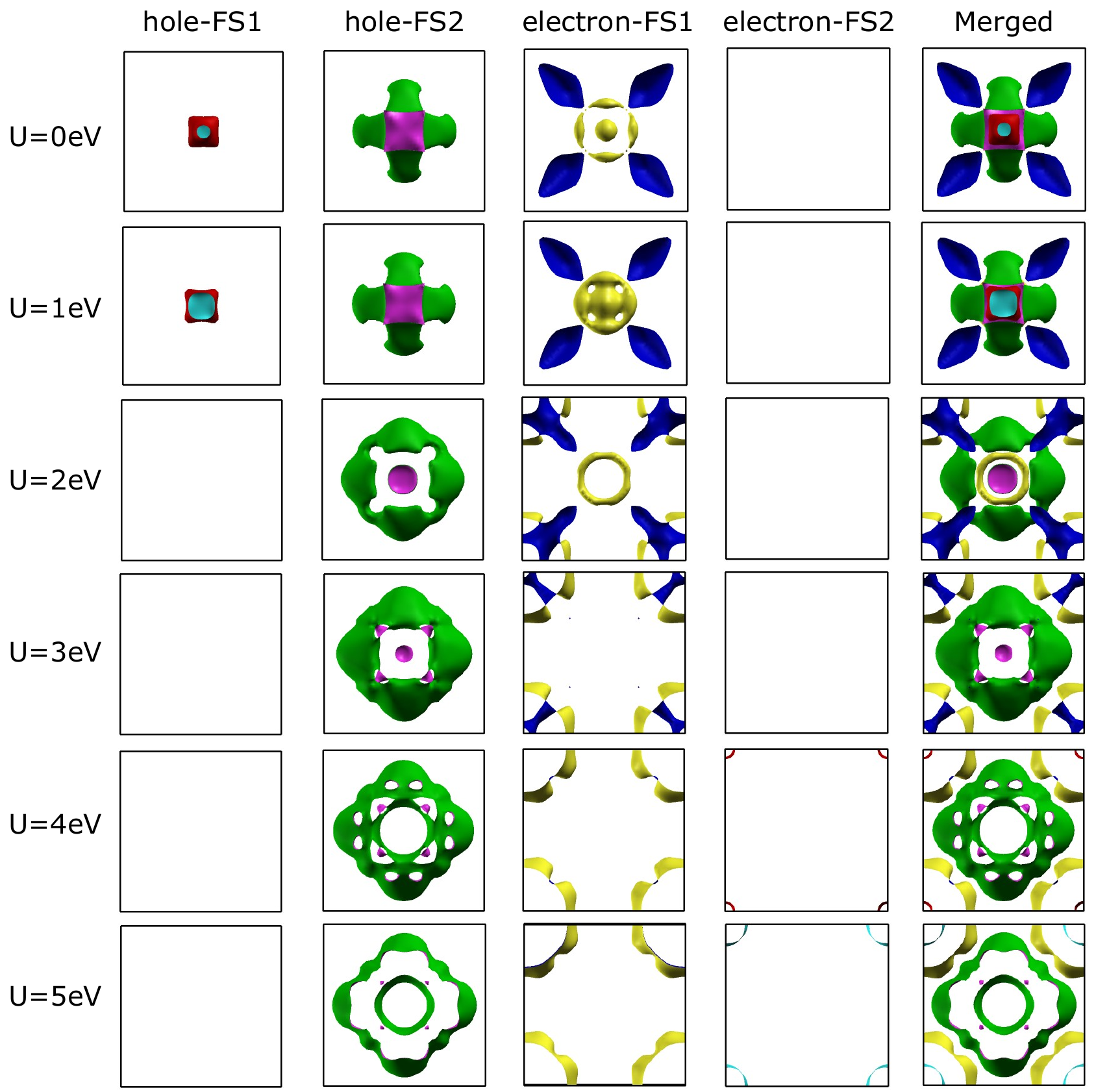}
\centering
\caption{FSs from the top-view by the DFT$+U$ calculation with various values of $U$.
\label{fig:FS_u0-5_1}}
\end{figure}

\begin{figure}[tbp]
\includegraphics[width=1.0\linewidth]{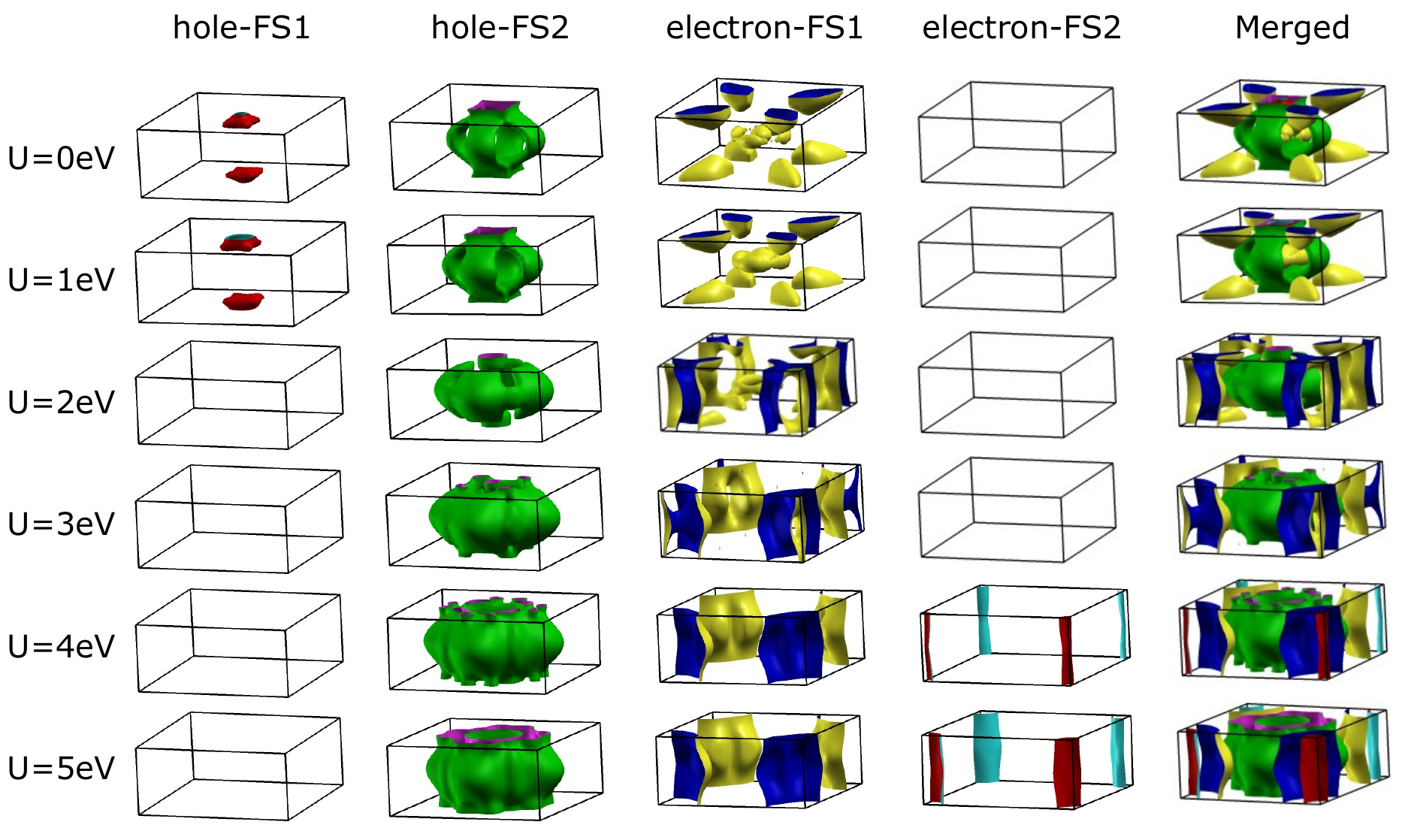}
\centering
\caption{Evolotion of FSs by the DFT$+U$ calculation with increasing the interaction parameter $U$.
\label{fig:FS_u0-5_2}}
\end{figure}

Figures \ref{fig:FS_u0-5_1} and \ref{fig:FS_u0-5_2} show the obtained four FSs, namely, the hole-FS1(2) and electron-FS1(2), in addition to the merged FSs. While the hole-FS1 disappears, the electron-FS2 appears by introducing $U$. 
With increasing $U$, the shape of hole-FS2 and electron-FS1 changes from a 3D shape to a quasi-2D one (see Fig.~\ref{fig:FS_u0-5_1}). 
Here we refer to the 3D (2D) nature of the FSs by their topology, whether they are closed (open) along the $k_z$ direction. The emergence of the 2D nature upon increasing the electron correlation is understood from the crystal structure and the evolution of the $c$-$f$ hybridization between different layers.
The DFT$+U$ result for $U=3$ eV is partly consistent with the renormalized band calculations \cite{Hafner2022Possible, Cavanagh2022Nonsymmorphic}: (i) the presence of the cylindrical hole-FS2 along the $\Gamma$-$Z$ line with a bulge in the basal plane $k_z=0$ and (ii) the cylindrical electron-FS1 at the zone boundary. 
However, a $\Gamma$-centered closed surface \cite{Hafner2022Possible, Cavanagh2022Nonsymmorphic} is absent in the present calculation.
It has been pointed out that the electron-FS1 accounts for 80$\%$ of the total density of states and may play a central role in stabilizing the superconducting state \cite{Cavanagh2022Nonsymmorphic}.
It is also intriguing that we observe disconnected electron- and hole-pockets with a 2D nature. This is similar to iron-based superconductors, which share the same space group symmetry. In the iron-based superconductors, several electric quadrupole orders have been observed, and fully gapped $s$-wave superconductivity emerges due to the electric quadrupole and antiferromagnetic spin fluctuations \cite{Mazin2008, Kuroki2008, Fernandes2010, Yanagi2010, Kontani2011, Onari2012}. 
Thus, the evolution of electronic structure due to the electron correlation may be essential for the superconductivity, and conversely, a future experimental study of gap structure will give an insight into the electronic structure.
The evolution of FSs as a function of $U$ leads to various topological Lifshitz transitions, which are crucially important for topological superconductivity as we discuss below. 

\subsection{Topological crystalline superconductivity}
\label{sec:TCSC}

Here, we discuss topological properties of superconductivity in CeRh$_2$As$_2$. First, we review the FS formulas for the topological invariants obtained in Ref.~\onlinecite{Nogaki2021topological}. We assume odd-parity superconductivity as ensured by the gap function of the PDW state~\cite{Yoshida2012Pair-density,Fischer2023Superconductivity}. We also assume that the time-reversal symmetry is broken in the superconducting phase because the PDW state is found in CeRh$_2$As$_2$ under the magnetic field parallel to the $c$ axis~\cite{Khim2021Field-induced}.
In this magnetic field, the unitary part of the point group is $C_{4h}$, and the superconducting order parameter is classified into $A_u$, $B_u$, and $E_u$ irreducible representations (IRs). 
The magnetic space group is represented by $P4/nm'm'$ ($\#$129.417), and importantly, the glide symmetry $G=\{\sigma_z |\bm a/2 + \bm b/2\}$ is preserved in this magnetic field. 
Hereafter, we focus on the glide-invariant planes $k_z=0, \pi$ and glide-odd superconductivity, namely, the $A_u$ and $B_u$ states corresponding to the PDW state. We can prove that (i) the glide operator anticommutes with the particle-hole operator $\{G, {\cal C}\}=0$, and (ii) the glide eigenvalues for the positive and negative sectors, $\mathfrak{g}^\pm = \pm i \exp[-i(k_x + k_y)/2]$, become pure imaginary in the restricted one-dimensional (1D) ${\bm k}$-space $k_x + k_y = 0$ ($\Gamma$-$M$ and $Z$-$A$ lines). From (i) and (ii), the particle-hole symmetry is closed in each glide sector. Thus, each glide sector on these lines is classified into 1D superconductors in class D.
Therefore, the corresponding topological invariants are given by the 1D $\mathbb{Z}_2$ numbers: 
\begin{equation}
 \nu_{k_z=0,\pi}^{\mathfrak{g}^\pm} = \frac{1}{\pi}\int^{\Gamma_2}_{\Gamma_1} d\,k_i \, \mathcal{A}^{\mathfrak{g}^\pm}_i(\bm k) 
  \hspace{5mm}({\rm mod} \,\,\, 2),
  \label{eq:z_2inv}
\end{equation}
where $\mathcal{A}_i^{\mathfrak{g}^\pm}(\bm k)$ is the $i$-th component of the Berry connection of each glide sector.
$\Gamma_1$ and $\Gamma_2$ denote the $\Gamma$ and $M$ points for $k_z=0$ and, analogously, the $Z$ and $A$ points for $k_z=\pi$.
We can identify whether $\nu_{k_z}^{\mathfrak{g}^{\pm}}$ is trivial or not from the number of Fermi surfaces \cite{Fu2010, Sato2010} by the fomula
\begin{equation}
\nu_{k_z}^{\mathfrak{g}^{\pm}} = \#{\rm FS}^{\pm}_{\Gamma_1\rightarrow\Gamma_2} \qquad({\rm mod}\ 2). \label{eq:Z2invariants}
\end{equation}

Next, we combine the FS formula with our DFT+$U$ calculations. We obtain $\nu^{\mathfrak{g}^{\pm}}_{k_z}$ in Table \ref{tab:TSC_1} from Eq.~\eqref{eq:Z2invariants} and the DFT+$U$ results in Figs.~\ref{fig:FS_u0-5_1} and \ref{fig:FS_u0-5_2}.
The nontrivial topological invariants $\nu^{\mathfrak{g}^{\pm}}_{k_z}$ are obtained for $U\leq3$ eV.
A topological transition occurs between $U=1$ and $2$ eV accompanied by the Lifshitz transition in the hole-FS1 pocket around the $Z$ point and the electron-FS1 pocket around the $\Gamma$ point. Thorough the topological transition, the invariants change from   $(\nu_0^{\mathfrak{g}^{\pm}}, \nu_\pi^{\mathfrak{g}^{\pm}})=(1, 0)$ to $(0, 1)$.
These two topological states predict similar surface Majorana fermions.
For a large $U \geq 4$ eV, the $\mathbb{Z}_2$ invariants are trivial in Table~\ref{tab:TSC_1}. 

In Table~\ref{tab:TSC_1}, we also show $C_{k_z=0, \pi}$ denoting the Chern number whose parity is related to the two Zak phases $\gamma$ and $\gamma'$ on the $k_x + k_y = 0$ and $k_x + k_y = \pi$ lines, respectively: $C_{k_z}=\gamma-\gamma'$ (mod $2$).
By definition, the Zak phase $\gamma$ on the $k_x + k_y = 0$ line is given by the sum of $\nu^{\mathfrak{g}^{\pm}}_{k_z}$.
Similar formulas to Eq.~\eqref{eq:Z2invariants} are valid for the Zak phase $\gamma'$ on the $k_x + k_y = \pi$ line and enforce $\gamma'$ to be zero since the occupation numbers are equivalent between the momenta $(\pi,0, k_z)$ and $(0,\pi, k_z)$ because of the fourfold rotation symmetry.
Thus, we obtain $C_{k_z}=\nu_{k_z}^{\mathfrak{g}^{+}}+\nu_{k_z}^{\mathfrak{g}^{-}}$ (mod 2), which is even according to Table \ref{tab:TSC_1}. This is consistent with the fact that the Chern number vanishes in the presence of time-reversal symmetry and our DFT+$U$ calculation is carried out at the zero magnetic field.

Here, we discuss the effect of the Zeeman splitting. The Zeeman field lifts the band degeneracy, and the $\mathbb{Z}_2$ invariants will change their values if the occupation numbers at the $\Gamma_1$ and $\Gamma_2$ points change. 
Since the $g$-factor in CeRh$_2$As$_2$ has not been determined, we estimate the Zeeman splitting with a value $g=10$ considering an enhancement due to an electron correlation effect~\cite{Nogaki2021topological}.
For $U=0$ and $4$ eV, we observe a shallow band across the Fermi level at the $Z$ point ($\sim4$ meV and $-7$ meV, respectively). It corresponds to the magnetic fields $7$ and $12$ T, which are lower than the upper critical field $H_{\rm c2}\sim14$ T.
Therefore, the Lifshitz transition may occur due to the magnetic field, and 
the Chern number $C_\pi$ and $\mathbb{Z}_2$ invariants $\nu^{\mathfrak{g}^{\pm}}_\pi$ may become nontrivial (Table~\ref{tab:TSC_2}). 
For different Chern numbers between $k_z =0$ and $\pi$, the Weyl superconductivity accompanied by the topological surface states is predicted~\cite{Nogaki2021topological,Yanase2016}.

To demonstrate the existence of Majorana states, 
we calculate $(\bar 110)$ surface states by using the {\it ab initio} Wannier model for $U=2.0$ eV. As shown in Sec.~S3~\cite{suppl}, Majorana states appear consistent with the analysis of topological invariants.

\begin{table}[tbp]
    \centering
    \caption{$\mathbb{Z}_2$ topological invariants, $\nu^{\mathfrak{g}^{+}}_{0}$, $\nu^{\mathfrak{g}^{-}}_{0}$, $\nu^{\mathfrak{g}^{+}}_{\pi}$,
    $\nu^{\mathfrak{g}^{-}}_{\pi}$, and Chern numbers $C_0$ and $C_\pi$ of CeRh$_2$As$_2$ in the PDW state for various values of $U$. 
    We assume that the occupation numbers at the $\Gamma_1$ and $\Gamma_2$ points are unchanged by the magnetic field.}
    \label{tab:TSC_1}
    \begin{tabular*}{1.0\columnwidth}{@{\extracolsep{\fill}}lccc} \hline\hline
     & $(\nu_0^{\mathfrak{g}^{+}}, \nu_0^{\mathfrak{g}^{-}}, C_0)$ & $(\nu_\pi^{\mathfrak{g}^{+}}, \nu_\pi^{\mathfrak{g}^{-}}, C_\pi)$ & Weyl SC \\ \hline
    $U=0$ eV & (1, 1, even) & (0, 0, even) & $\times$\\
    $U=1$ eV & (1, 1, even) & (0, 0, even) & $\times$\\
    $U=2$ eV & (0, 0, even) & (1, 1, even) & $\times$\\
    $U=3$ eV & (0, 0, even) & (1, 1, even) & $\times$\\
    $U=4$ eV & (0, 0, even) & (0, 0, even) & $\times$\\
    $U=5$ eV & (0, 0, even) & (0, 0, even) & $\times$\\ \hline\hline
    \end{tabular*}
\end{table}

\begin{table}[tbp]
    \centering
    \caption{Same as Table~\ref{tab:TSC_1}, but we assume the high magnetic field region. The occupation number at the $Z$ point can change due to the Zeeman splitting because the band can be sufficiently flat at the Fermi level near the $Z$ point. Difference in the Chern numbers at $k_z=0$ and $\pi$ indicates Weyl superconductivity. We will not discuss the Weyl superconductivity in the case of $C_{k_z} \in 2\mathbb{Z}$ since the integer part of $C_{k_z}/2$ depends on the detailed properties of the gap function.}
    \label{tab:TSC_2}
    \begin{tabular*}{1.0\columnwidth}{@{\extracolsep{\fill}}lccc} \hline\hline
    High field & $(\nu_0^{\mathfrak{g}^{+}}, \nu_0^{\mathfrak{g}^{-}}, C_0)$ & $(\nu_\pi^{\mathfrak{g}^{+}}, \nu_\pi^{\mathfrak{g}^{-}}, C_\pi)$ & Weyl SC \\ \hline
    $U=0$ eV & (1, 1, even) & (1, 0, odd) or (0, 1, odd) & $\circ$\\
    $U=1$ eV & (1, 1, even) & (0, 0, even) & \\
    $U=2$ eV & (0, 0, even) & (1, 1, even) & \\
    $U=3$ eV & (0, 0, even) & (1, 1, even) & \\
    $U=4$ eV & (0, 0, even) & (1, 0, odd) or (0, 1, odd) & $\circ$\\
    $U=5$ eV & (0, 0, even) & (0, 0, even) & \\ \hline\hline
    \end{tabular*}
\end{table}

\subsection{Superconducting gap nodes}

In this section, we elucidate the gap structure in the high-field superconducting state of CeRh$_2$As$_2$. 
We first give an ordinary classification theory on superconducting gap nodes based on the point group symmetry~\cite{Sigrist-Ueda}. In the absence of the magnetic field, superconducting order parameters are classified into the IRs of the $D_{4h}$ point group. 
In the presence of the magnetic field, the unitary part of the point group is reduced to $C_{4h}$. Thus, we can deduce nodes from the basis function of 
$D_{4h}$ and the compatibility relation to $C_{4h}$.
The gap function is given by $\hat{\Delta}(\boldsymbol{k}) = \left( \psi_0(\boldsymbol{k}) \hat{\sigma}_{0} i\hat{\sigma}_{y} \otimes \hat{\tau}_{z} + \boldsymbol{d}(\boldsymbol{k}) \cdot \hat{\boldsymbol{\sigma}}i\hat{\sigma}_{y} \otimes \hat{\tau}_{0} \right)$,
where $\sigma_\mu$ and $\tau_\mu$ are Pauli matrices for spin and sublattice space.
We ignored the orbital degree of freedom. 
We limit ourselves to the case in which superconductivity is induced by the intra-sublattice Cooper pairs of Ce $4f$ electrons.
The corresponding basis functions are
\begin{align}
    \boldsymbol{d}_{A_{u}}(\boldsymbol{k}) &= a_1(k_x \hat{{\bm x}} + k_y \hat{{\bm y}}) + a_2 k_z \hat{{\bm z}} + a_3(k_y \hat{\bm{x}} - k_x \hat{\bm{y}}),\\
    \psi_{A_{u}}(\boldsymbol{k}) &= a_4 [1+k_xk_y(k_x^2-k_y^2)],
\end{align}
and
\begin{align}
    {\bm d}_{B_{u}}(\boldsymbol{k})& = b_1(k_x \hat{\bm{x}} - k_y \hat{\bm{y}}) + b_2 (k_y \hat{\bm{x}} + k_x \hat{\bm{y}}), \\
    \psi_{B_{u}}(\boldsymbol{k}) &= b_3 k_xk_y + b_4(k_x^2-k_y^2).
\end{align}
Considering the basis function, we observe that the $B_u$ pairing state has point nodes on the $k_z$ axis.
The $A_u$ pairing state may appear to be gapped since the staggered $s$-wave pairing ($\psi_{A_{u}}(\boldsymbol{k})=a_4$) remains finite in the whole ${\bm k}$ space, but we will see that also the $A_u$ pairing state can have point nodes on the $k_z$ axis.
For the $E_u$ pairing state, the gap function is
\begin{align}
    {\bm d}_{E^{1,2}_{u}}(\boldsymbol{k}) & = c_1(k_x \pm ik_y)\hat{\bm z} + c_2 k_z(\hat{\bm x} \pm i\hat{\bm y}), \\
    \psi_{E^{1,2}_{u}}(\boldsymbol{k}) & = c_3 (k_xk_z \pm i k_yk_z), \label{d_Eu}
\end{align}
where the upper (lower) sign corresponds to the $E^{1}_{u}$ ($E^{2}_{u}$) state. 
Note that the magnetic field lifts the degeneracy of the $E_{u}$ states. 
Under the high magnetic field along the $c$ axis, the $d$-vector parallel to the magnetic field ($\hat{\bm z}$-component) is suppressed. In such a case, we expect a (pseudo-) line node on the $k_z=0$ plane.

Recently, it has been shown that the superconducting gap structures at given high-symmetry points in $\bm k$ space are precisely predicted by the topological classification \cite{Kobayashi2014, Kobayashi2016, Kobayashi2018, Sumita2019,Ono2022Symmetry-Based}, in which the gapless points are characterized by topological invariants.
Especially, gap nodes protected by the nonsymmorphic space group symmetry and the total angular momentum of electrons are successfully found, consistent with previous group theoretical classifications~\cite{Yarzhemsky1992, Norman1995, Micklitz2009, Nomoto2017, Micklitz2017_PRL, Sumita2017, Sumita2018}.
Here, we classify the gap structure on the three different lines: (i) $k_x+k_y=0$ lines with $k_z=0$ and $\pi$, and (ii) $k_x=k_y=0$ line ($k_z$ axis).
The method of classification is given in the Supplemental Material~\cite{suppl}.

Summary of classification of gap nodes is shown in Table~\ref{tab:Gap_Class}.
For case (i), the $A_u$ IR in $C_{2h}$ point group corresponds to the PDW state of $A_u$ and $B_u$ representations in $C_{4h}$.
The classification with index 0 ($\mathbb{Z}$) indicates that the gap {\it opens} ({\it closes}) by the superconducting pair wave function when FSs in the normal state exist on the lines.
In Table \ref{tab:Gap_Topo} we summarize the excitation nodes for each pairing symmetry and FSs. 
For example, the $A_u$ pairing state has a point node on the $k_z$ axis for $U\leq4$ eV (we again assumed the high magnetic field), while it is fully gapped for $U\geq5$ eV. 
Table~\ref{tab:Gap_Class} is consistent with the result in Ref.~\onlinecite{Ono2022Symmetry-Based}.

Our analysis is based on the symmetry, and then it is assumed that the time-reversal symmetry is entirely broken. 
It corresponds to the situation that the magnetic field is sufficiently strong, and the $d$-vector parallel to the magnetic field is completely suppressed.
For the $E^{1,2}_{u}$ states, we expect a line node on the $k_z=0$ plane from Eq.~\eqref{d_Eu}, consistent with the nodes along the $k_x+k_y=0$ lines. 
Although the topological classification does not give us any information on off-axis ${\bm k}$-points, the node on the $k_x+k_y=0$ line should be interpreted as a part of the line node on the $k_z=0$ plane.

It should be noticed that the TCSC discussed in Sec.~\ref{sec:TCSC} is stable against excitation nodes.
The 1D topological invariants on the $k_x+k_y=0$ lines are well-defined in any values of $U$, because the superconducting gap opens on these lines in the $A_u$ and $B_u$ superconducting states expected in CeRh$_2$As$_2$ at high magnetic fields. The corresponding Majorana surface states appear at the $(\bar 110)$ surface preserving the glide symmetry when the topological invariants in Tables~\ref{tab:TSC_1} and \ref{tab:TSC_2} are nontrivial. 

\begin{table}[tbp]
    \centering    \caption{Classification of superconducting gap structures for all odd-parity IRs under the magnetic field parallel to the $c$ axis. Results of the Wigner criteria and orthogonality test, effective Altland-Zirnbauer (EAZ) class, and topological classification~\cite{suppl} on the high symmetry lines (i) $k_x+k_y=0$ with $k_z=0$, $\pi$ and (ii) $k_x=k_y=0$ ($k_z$ axis) are shown. For case (i), the $A_u$ IR in $C_{2h}$ point group corresponds to the PDW state expected in CeRh$_2$As$_2$.}
    \label{tab:Gap_Class}
    \begin{tabular*}{1.0\columnwidth}{@{\extracolsep{\fill}}lccccc} \hline\hline
    \multicolumn{6}{l}{(i) $k_x+k_y=0$, $k_z=0$ and $\pi$}  \\
    \hline
    IR of $C_{2h}$ & $W^{\mathfrak{T}}$ & $W^{\mathfrak{C}}$ & $W^{\Gamma}$ & EAZ Class & Classification \\
    $A_u$ & 1 & $-1$ & 1 & CI & 0 \\
    $B_u$ & 1 & 0 & 0 & AI & $\mathbb{Z}$ \\
    \hline
    \multicolumn{6}{l}{(ii) $k_x=k_y=0$ ($k_z$ axis)}  \\
    \hline
    IR of $C_{4h}$ & $W^{\mathfrak{T}}$ & $W^{\mathfrak{C}}$ & $W^{\Gamma}$ & EAZ Class & Classification \\
    $A_u$ & 1 & 0 & 0 & AI & $\mathbb{Z}$ \\
    $B_u$ & 1 & 0 & 0 & AI & $\mathbb{Z}$ \\
    $E^1_u$ & 1 & 0 & 0 & AI & $\mathbb{Z}$ \\
    $E^2_u$ & 1 & $-1$ & 1 & CI & 0 \\
 \hline\hline
    \end{tabular*}
\end{table}

\begin{table}[tbp]
    \centering
    \caption{Excitation nodes and topological indices for odd-parity pairing states. A node of the superconducting gap appears on the $k_z$ axis when FSs intersect the axis: $U\leq4$ eV with the high magnetic field, while gapped otherwise.}
    \label{tab:Gap_Topo}
    \begin{tabular*}{1.0\columnwidth}{@{\extracolsep{\fill}}lccc} \hline\hline
    & $k_x+k_y=0$ lines & $k_z$ axis & Topological index \\ \hline
    $A_u$ & gap & node/gap & $(\nu_{k_z}^{\mathfrak{g}^{+}}, \nu_{k_z}^{\mathfrak{g}^{-}}, C_{k_z})$ \\
    $B_u$ & gap & node/gap & $(\nu_{k_z}^{\mathfrak{g}^{+}}, \nu_{k_z}^{\mathfrak{g}^{-}}, C_{k_z})$ \\
    $E^1_u$ & node & node/gap & $C_{k_z}$ \\
    $E^2_u$ & node & gap & $C_{k_z}$ \\
 \hline\hline
    \end{tabular*}
\end{table}

\section{Summary}
In this Letter we have investigated the electronic state in CeRh$_2$As$_2$ by DFT$+U$ method and shown how the electron correlation changes the band structure, Fermi surface, and topological properties of superconductivity.
A strong hybridization between Rh and Ce atoms at the Van Hove singularity and heavy electron pockets near the Brillouin zone corner are highlighted.
The evolution of the Fermi surface as a function of $U$ revealed a crossover from a 3D electron system to a quasi-2D one. 
A weak interlayer coupling favoring the 2D electronic structure and FSs near the zone corner of nonsymmorphic crystals are favorable features for the two superconducting phases with even- and odd-parity~\cite{Yoshida2012Pair-density,sumita2016,Cavanagh2022Nonsymmorphic}. Thus, the correlation-induced FS evolution supports the PDW scenario for CeRh$_2$As$_2$.

The similarities of CeRh$_2$As$_2$ and iron-based superconductors were discussed. The hole pockets at the zone center and the electron pockets at the zone corner are similar to the iron-based superconductors.
While the nesting vector for the magnetic and quadrupole orders has been discussed~\cite{Kibune2022CeRh2As2, Hafner2022Possible}, we expect the finite $\bm q$ nesting vector from our results: it would be ${\bm q} =(\pi, \pi, 0)$ if the electron correlation is strong enough. 

The topological $\mathbb{Z}_2$ invariants for TCSC are evaluated from the Fermi surface formula, and the odd-parity superconducting states in the $A_u$ and $B_u$ representations are identified as the TCSC unless $U$ is significantly large.
Superconducting gap structures under the magnetic field are predicted by classification of symmetry and topology. 
The superconducting gap protecting the $\mathbb{Z}_2$ invariants are found to be robust, and
Majorana surface states appear at the $(\bar{1}10)$ surface preserving the glide symmetry. 
Because the high-field superconducting phase of CeRh$_2$As$_2$ is expected to be of the $A_u$ or $B_u$ representation, TCSC is predicted for a wide range of the parameter $U$.
Although the band dispersion along the high symmetry line agrees with the ARPES experiment~\cite{Chen2023Coexistence}, the presence or absence of a small FS has not been investigated experimentally.
Our calculations enable us to determine the topology in the PDW state from the normal electronic band structure, and thus, the observation of whole FSs is desirable for experimental verification.

In this Letter, we have ignored the effects of normal state order on superconductivity. This assumption is supported by the decoupling of the PDW state from the quadrupole order~\cite{Semeniuk2023Decoupling} and the magnetic order~\cite{Ogata2023Parity}. Because the normal state order is suppressed in the PDW state at high magnetic fields or pressures, our treatment is justified there. Even when the glide symmetry is broken by the quadrupole order in the coexisting region, a surface state reminiscent of the Majorana fermion is expected to appear.


\begin{acknowledgments}
We appreciate helpful discussions with Y. Huang, B. Lv, X. Chen, E. Hassinger, K. Ishida, and M. H. Fischer.
This work was supported by JSPS KAKENHI (Grants No. JP21K18145, JP22H04933, JP22H01181, JP22KJ1716, JP23K17353, JP24K21530, and JP24H00007) by the Swiss National Science Foundation (SNSF) Division II (No. 184739).
\end{acknowledgments}


\begin{thebibliography}{93}%
    \makeatletter
    \providecommand \@ifxundefined [1]{%
     \@ifx{#1\undefined}
    }%
    \providecommand \@ifnum [1]{%
     \ifnum #1\expandafter \@firstoftwo
     \else \expandafter \@secondoftwo
     \fi
    }%
    \providecommand \@ifx [1]{%
     \ifx #1\expandafter \@firstoftwo
     \else \expandafter \@secondoftwo
     \fi
    }%
    \providecommand \natexlab [1]{#1}%
    \providecommand \enquote  [1]{``#1''}%
    \providecommand \bibnamefont  [1]{#1}%
    \providecommand \bibfnamefont [1]{#1}%
    \providecommand \citenamefont [1]{#1}%
    \providecommand \href@noop [0]{\@secondoftwo}%
    \providecommand \href [0]{\begingroup \@sanitize@url \@href}%
    \providecommand \@href[1]{\@@startlink{#1}\@@href}%
    \providecommand \@@href[1]{\endgroup#1\@@endlink}%
    \providecommand \@sanitize@url [0]{\catcode `\\12\catcode `\$12\catcode `\&12\catcode `\#12\catcode `\^12\catcode `\_12\catcode `\%12\relax}%
    \providecommand \@@startlink[1]{}%
    \providecommand \@@endlink[0]{}%
    \providecommand \url  [0]{\begingroup\@sanitize@url \@url }%
    \providecommand \@url [1]{\endgroup\@href {#1}{\urlprefix }}%
    \providecommand \urlprefix  [0]{URL }%
    \providecommand \Eprint [0]{\href }%
    \providecommand \doibase [0]{http://dx.doi.org/}%
    \providecommand \selectlanguage [0]{\@gobble}%
    \providecommand \bibinfo  [0]{\@secondoftwo}%
    \providecommand \bibfield  [0]{\@secondoftwo}%
    \providecommand \translation [1]{[#1]}%
    \providecommand \BibitemOpen [0]{}%
    \providecommand \bibitemStop [0]{}%
    \providecommand \bibitemNoStop [0]{.\EOS\space}%
    \providecommand \EOS [0]{\spacefactor3000\relax}%
    \providecommand \BibitemShut  [1]{\csname bibitem#1\endcsname}%
    \let\auto@bib@innerbib\@empty
    \bibitem [{\citenamefont {Khim}\ \emph {et~al.}(2021)\citenamefont {Khim}, \citenamefont {Landaeta}, \citenamefont {Banda}, \citenamefont {Bannor}, \citenamefont {Brando}, \citenamefont {Brydon}, \citenamefont {Hafner}, \citenamefont {K{\"u}chler}, \citenamefont {Cardoso-Gil}, \citenamefont {Stockert}, \citenamefont {Mackenzie}, \citenamefont {Agterberg}, \citenamefont {Geibel},\ and\ \citenamefont {Hassinger}}]{Khim2021Field-induced}%
      \BibitemOpen
      \bibfield  {author} {\bibinfo {author} {\bibfnamefont {S.}~\bibnamefont {Khim}}, \bibinfo {author} {\bibfnamefont {J.~F.}\ \bibnamefont {Landaeta}}, \bibinfo {author} {\bibfnamefont {J.}~\bibnamefont {Banda}}, \bibinfo {author} {\bibfnamefont {N.}~\bibnamefont {Bannor}}, \bibinfo {author} {\bibfnamefont {M.}~\bibnamefont {Brando}}, \bibinfo {author} {\bibfnamefont {P.~M.~R.}\ \bibnamefont {Brydon}}, \bibinfo {author} {\bibfnamefont {D.}~\bibnamefont {Hafner}}, \bibinfo {author} {\bibfnamefont {R.}~\bibnamefont {K{\"u}chler}}, \bibinfo {author} {\bibfnamefont {R.}~\bibnamefont {Cardoso-Gil}}, \bibinfo {author} {\bibfnamefont {U.}~\bibnamefont {Stockert}}, \bibinfo {author} {\bibfnamefont {A.~P.}\ \bibnamefont {Mackenzie}}, \bibinfo {author} {\bibfnamefont {D.~F.}\ \bibnamefont {Agterberg}}, \bibinfo {author} {\bibfnamefont {C.}~\bibnamefont {Geibel}}, \ and\ \bibinfo {author} {\bibfnamefont {E.}~\bibnamefont {Hassinger}},\ }\href {\doibase DOI: 10.1126/science.abe7518} {\bibfield  {journal} {\bibinfo
      {journal} {Science}\ }\textbf {\bibinfo {volume} {373}},\ \bibinfo {pages} {1012} (\bibinfo {year} {2021})}\BibitemShut {NoStop}%
    \bibitem [{\citenamefont {Landaeta}\ \emph {et~al.}(2022)\citenamefont {Landaeta}, \citenamefont {Khanenko}, \citenamefont {Cavanagh}, \citenamefont {Geibel}, \citenamefont {Khim}, \citenamefont {Mishra}, \citenamefont {Sheikin}, \citenamefont {Brydon}, \citenamefont {Agterberg}, \citenamefont {Brando},\ and\ \citenamefont {Hassinger}}]{Landaeta2022Field-Angle}%
      \BibitemOpen
      \bibfield  {author} {\bibinfo {author} {\bibfnamefont {J.~F.}\ \bibnamefont {Landaeta}}, \bibinfo {author} {\bibfnamefont {P.}~\bibnamefont {Khanenko}}, \bibinfo {author} {\bibfnamefont {D.~C.}\ \bibnamefont {Cavanagh}}, \bibinfo {author} {\bibfnamefont {C.}~\bibnamefont {Geibel}}, \bibinfo {author} {\bibfnamefont {S.}~\bibnamefont {Khim}}, \bibinfo {author} {\bibfnamefont {S.}~\bibnamefont {Mishra}}, \bibinfo {author} {\bibfnamefont {I.}~\bibnamefont {Sheikin}}, \bibinfo {author} {\bibfnamefont {P.~M.~R.}\ \bibnamefont {Brydon}}, \bibinfo {author} {\bibfnamefont {D.~F.}\ \bibnamefont {Agterberg}}, \bibinfo {author} {\bibfnamefont {M.}~\bibnamefont {Brando}}, \ and\ \bibinfo {author} {\bibfnamefont {E.}~\bibnamefont {Hassinger}},\ }\href {\doibase 10.1103/PhysRevX.12.031001} {\bibfield  {journal} {\bibinfo  {journal} {Phys. Rev. X}\ }\textbf {\bibinfo {volume} {12}},\ \bibinfo {pages} {031001} (\bibinfo {year} {2022})}\BibitemShut {NoStop}%
    \bibitem [{\citenamefont {Yoshida}\ \emph {et~al.}(2012)\citenamefont {Yoshida}, \citenamefont {Sigrist},\ and\ \citenamefont {Yanase}}]{Yoshida2012Pair-density}%
      \BibitemOpen
      \bibfield  {author} {\bibinfo {author} {\bibfnamefont {T.}~\bibnamefont {Yoshida}}, \bibinfo {author} {\bibfnamefont {M.}~\bibnamefont {Sigrist}}, \ and\ \bibinfo {author} {\bibfnamefont {Y.}~\bibnamefont {Yanase}},\ }\href {\doibase 10.1103/PhysRevB.86.134514} {\bibfield  {journal} {\bibinfo  {journal} {Phys. Rev. B}\ }\textbf {\bibinfo {volume} {86}},\ \bibinfo {pages} {134514} (\bibinfo {year} {2012})}\BibitemShut {NoStop}%
    \bibitem [{\citenamefont {Fischer}\ \emph {et~al.}(2023)\citenamefont {Fischer}, \citenamefont {Sigrist}, \citenamefont {Agterberg},\ and\ \citenamefont {Yanase}}]{Fischer2023Superconductivity}%
      \BibitemOpen
      \bibfield  {author} {\bibinfo {author} {\bibfnamefont {M.~H.}\ \bibnamefont {Fischer}}, \bibinfo {author} {\bibfnamefont {M.}~\bibnamefont {Sigrist}}, \bibinfo {author} {\bibfnamefont {D.~F.}\ \bibnamefont {Agterberg}}, \ and\ \bibinfo {author} {\bibfnamefont {Y.}~\bibnamefont {Yanase}},\ }\href {\doibase https://doi.org/10.1146/annurev-conmatphys-040521-042511} {\bibfield  {journal} {\bibinfo  {journal} {Annual Review of Condensed Matter Physics}\ }\textbf {\bibinfo {volume} {14}},\ \bibinfo {pages} {153} (\bibinfo {year} {2023})}\BibitemShut {NoStop}%
    \bibitem [{\citenamefont {Ogata}\ \emph {et~al.}(2023{\natexlab{a}})\citenamefont {Ogata}, \citenamefont {Kitagawa}, \citenamefont {Kinjo}, \citenamefont {Ishida}, \citenamefont {Brando}, \citenamefont {Hassinger}, \citenamefont {Geibel},\ and\ \citenamefont {Khim}}]{Ogata2023Parity}%
      \BibitemOpen
      \bibfield  {author} {\bibinfo {author} {\bibfnamefont {S.}~\bibnamefont {Ogata}}, \bibinfo {author} {\bibfnamefont {S.}~\bibnamefont {Kitagawa}}, \bibinfo {author} {\bibfnamefont {K.}~\bibnamefont {Kinjo}}, \bibinfo {author} {\bibfnamefont {K.}~\bibnamefont {Ishida}}, \bibinfo {author} {\bibfnamefont {M.}~\bibnamefont {Brando}}, \bibinfo {author} {\bibfnamefont {E.}~\bibnamefont {Hassinger}}, \bibinfo {author} {\bibfnamefont {C.}~\bibnamefont {Geibel}}, \ and\ \bibinfo {author} {\bibfnamefont {S.}~\bibnamefont {Khim}},\ }\href {\doibase 10.1103/PhysRevLett.130.166001} {\bibfield  {journal} {\bibinfo  {journal} {Phys. Rev. Lett.}\ }\textbf {\bibinfo {volume} {130}},\ \bibinfo {pages} {166001} (\bibinfo {year} {2023}{\natexlab{a}})}\BibitemShut {NoStop}%
    \bibitem [{\citenamefont {Hafner}\ \emph {et~al.}(2022)\citenamefont {Hafner}, \citenamefont {Khanenko}, \citenamefont {Eljaouhari}, \citenamefont {K\"uchler}, \citenamefont {Banda}, \citenamefont {Bannor}, \citenamefont {L\"uhmann}, \citenamefont {Landaeta}, \citenamefont {Mishra}, \citenamefont {Sheikin}, \citenamefont {Hassinger}, \citenamefont {Khim}, \citenamefont {Geibel}, \citenamefont {Zwicknagl},\ and\ \citenamefont {Brando}}]{Hafner2022Possible}%
      \BibitemOpen
      \bibfield  {author} {\bibinfo {author} {\bibfnamefont {D.}~\bibnamefont {Hafner}}, \bibinfo {author} {\bibfnamefont {P.}~\bibnamefont {Khanenko}}, \bibinfo {author} {\bibfnamefont {E.-O.}\ \bibnamefont {Eljaouhari}}, \bibinfo {author} {\bibfnamefont {R.}~\bibnamefont {K\"uchler}}, \bibinfo {author} {\bibfnamefont {J.}~\bibnamefont {Banda}}, \bibinfo {author} {\bibfnamefont {N.}~\bibnamefont {Bannor}}, \bibinfo {author} {\bibfnamefont {T.}~\bibnamefont {L\"uhmann}}, \bibinfo {author} {\bibfnamefont {J.~F.}\ \bibnamefont {Landaeta}}, \bibinfo {author} {\bibfnamefont {S.}~\bibnamefont {Mishra}}, \bibinfo {author} {\bibfnamefont {I.}~\bibnamefont {Sheikin}}, \bibinfo {author} {\bibfnamefont {E.}~\bibnamefont {Hassinger}}, \bibinfo {author} {\bibfnamefont {S.}~\bibnamefont {Khim}}, \bibinfo {author} {\bibfnamefont {C.}~\bibnamefont {Geibel}}, \bibinfo {author} {\bibfnamefont {G.}~\bibnamefont {Zwicknagl}}, \ and\ \bibinfo {author} {\bibfnamefont {M.}~\bibnamefont {Brando}},\ }\href {\doibase
      10.1103/PhysRevX.12.011023} {\bibfield  {journal} {\bibinfo  {journal} {Phys. Rev. X}\ }\textbf {\bibinfo {volume} {12}},\ \bibinfo {pages} {011023} (\bibinfo {year} {2022})}\BibitemShut {NoStop}%
    \bibitem [{\citenamefont {Mishra}\ \emph {et~al.}(2022)\citenamefont {Mishra}, \citenamefont {Liu}, \citenamefont {Bauer}, \citenamefont {Ronning},\ and\ \citenamefont {Thomas}}]{Mishra2022Anisotropic}%
      \BibitemOpen
      \bibfield  {author} {\bibinfo {author} {\bibfnamefont {S.}~\bibnamefont {Mishra}}, \bibinfo {author} {\bibfnamefont {Y.}~\bibnamefont {Liu}}, \bibinfo {author} {\bibfnamefont {E.~D.}\ \bibnamefont {Bauer}}, \bibinfo {author} {\bibfnamefont {F.}~\bibnamefont {Ronning}}, \ and\ \bibinfo {author} {\bibfnamefont {S.~M.}\ \bibnamefont {Thomas}},\ }\href {\doibase 10.1103/PhysRevB.106.L140502} {\bibfield  {journal} {\bibinfo  {journal} {Phys. Rev. B}\ }\textbf {\bibinfo {volume} {106}},\ \bibinfo {pages} {L140502} (\bibinfo {year} {2022})}\BibitemShut {NoStop}%
    \bibitem [{\citenamefont {Semeniuk}\ \emph {et~al.}(2023)\citenamefont {Semeniuk}, \citenamefont {Hafner}, \citenamefont {Khanenko}, \citenamefont {L\"uhmann}, \citenamefont {Banda}, \citenamefont {Landaeta}, \citenamefont {Geibel}, \citenamefont {Khim}, \citenamefont {Hassinger},\ and\ \citenamefont {Brando}}]{Semeniuk2023Decoupling}%
      \BibitemOpen
      \bibfield  {author} {\bibinfo {author} {\bibfnamefont {K.}~\bibnamefont {Semeniuk}}, \bibinfo {author} {\bibfnamefont {D.}~\bibnamefont {Hafner}}, \bibinfo {author} {\bibfnamefont {P.}~\bibnamefont {Khanenko}}, \bibinfo {author} {\bibfnamefont {T.}~\bibnamefont {L\"uhmann}}, \bibinfo {author} {\bibfnamefont {J.}~\bibnamefont {Banda}}, \bibinfo {author} {\bibfnamefont {J.~F.}\ \bibnamefont {Landaeta}}, \bibinfo {author} {\bibfnamefont {C.}~\bibnamefont {Geibel}}, \bibinfo {author} {\bibfnamefont {S.}~\bibnamefont {Khim}}, \bibinfo {author} {\bibfnamefont {E.}~\bibnamefont {Hassinger}}, \ and\ \bibinfo {author} {\bibfnamefont {M.}~\bibnamefont {Brando}},\ }\href {\doibase 10.1103/PhysRevB.107.L220504} {\bibfield  {journal} {\bibinfo  {journal} {Phys. Rev. B}\ }\textbf {\bibinfo {volume} {107}},\ \bibinfo {pages} {L220504} (\bibinfo {year} {2023})}\BibitemShut {NoStop}%
    \bibitem [{\citenamefont {Kibune}\ \emph {et~al.}(2022)\citenamefont {Kibune}, \citenamefont {Kitagawa}, \citenamefont {Kinjo}, \citenamefont {Ogata}, \citenamefont {Manago}, \citenamefont {Taniguchi}, \citenamefont {Ishida}, \citenamefont {Brando}, \citenamefont {Hassinger}, \citenamefont {Rosner}, \citenamefont {Geibel},\ and\ \citenamefont {Khim}}]{Kibune2022CeRh2As2}%
      \BibitemOpen
      \bibfield  {author} {\bibinfo {author} {\bibfnamefont {M.}~\bibnamefont {Kibune}}, \bibinfo {author} {\bibfnamefont {S.}~\bibnamefont {Kitagawa}}, \bibinfo {author} {\bibfnamefont {K.}~\bibnamefont {Kinjo}}, \bibinfo {author} {\bibfnamefont {S.}~\bibnamefont {Ogata}}, \bibinfo {author} {\bibfnamefont {M.}~\bibnamefont {Manago}}, \bibinfo {author} {\bibfnamefont {T.}~\bibnamefont {Taniguchi}}, \bibinfo {author} {\bibfnamefont {K.}~\bibnamefont {Ishida}}, \bibinfo {author} {\bibfnamefont {M.}~\bibnamefont {Brando}}, \bibinfo {author} {\bibfnamefont {E.}~\bibnamefont {Hassinger}}, \bibinfo {author} {\bibfnamefont {H.}~\bibnamefont {Rosner}}, \bibinfo {author} {\bibfnamefont {C.}~\bibnamefont {Geibel}}, \ and\ \bibinfo {author} {\bibfnamefont {S.}~\bibnamefont {Khim}},\ }\href {\doibase 10.1103/PhysRevLett.128.057002} {\bibfield  {journal} {\bibinfo  {journal} {Phys. Rev. Lett.}\ }\textbf {\bibinfo {volume} {128}},\ \bibinfo {pages} {057002} (\bibinfo {year} {2022})}\BibitemShut {NoStop}%
    \bibitem [{\citenamefont {Kitagawa}\ \emph {et~al.}(2022)\citenamefont {Kitagawa}, \citenamefont {Kibune}, \citenamefont {Kinjo}, \citenamefont {Manago}, \citenamefont {Taniguchi}, \citenamefont {Ishida}, \citenamefont {Brando}, \citenamefont {Hassinger}, \citenamefont {Geibel},\ and\ \citenamefont {Khim}}]{Kitagawa2022Two-Dimensional}%
      \BibitemOpen
      \bibfield  {author} {\bibinfo {author} {\bibfnamefont {S.}~\bibnamefont {Kitagawa}}, \bibinfo {author} {\bibfnamefont {M.}~\bibnamefont {Kibune}}, \bibinfo {author} {\bibfnamefont {K.}~\bibnamefont {Kinjo}}, \bibinfo {author} {\bibfnamefont {M.}~\bibnamefont {Manago}}, \bibinfo {author} {\bibfnamefont {T.}~\bibnamefont {Taniguchi}}, \bibinfo {author} {\bibfnamefont {K.}~\bibnamefont {Ishida}}, \bibinfo {author} {\bibfnamefont {M.}~\bibnamefont {Brando}}, \bibinfo {author} {\bibfnamefont {E.}~\bibnamefont {Hassinger}}, \bibinfo {author} {\bibfnamefont {C.}~\bibnamefont {Geibel}}, \ and\ \bibinfo {author} {\bibfnamefont {S.}~\bibnamefont {Khim}},\ }\href {\doibase 10.7566/JPSJ.91.043702} {\bibfield  {journal} {\bibinfo  {journal} {J. Phys. Soc. Jpn.}\ }\textbf {\bibinfo {volume} {91}},\ \bibinfo {pages} {043702} (\bibinfo {year} {2022})}\BibitemShut {NoStop}%
    \bibitem [{\citenamefont {Ogata}\ \emph {et~al.}(2023{\natexlab{b}})\citenamefont {Ogata}, \citenamefont {Kitagawa}, \citenamefont {Kinjo}, \citenamefont {Ishida}, \citenamefont {Brando}, \citenamefont {Hassinger}, \citenamefont {Geibel},\ and\ \citenamefont {Khim}}]{Ogata2023}%
      \BibitemOpen
      \bibfield  {author} {\bibinfo {author} {\bibfnamefont {S.}~\bibnamefont {Ogata}}, \bibinfo {author} {\bibfnamefont {S.}~\bibnamefont {Kitagawa}}, \bibinfo {author} {\bibfnamefont {K.}~\bibnamefont {Kinjo}}, \bibinfo {author} {\bibfnamefont {K.}~\bibnamefont {Ishida}}, \bibinfo {author} {\bibfnamefont {M.}~\bibnamefont {Brando}}, \bibinfo {author} {\bibfnamefont {E.}~\bibnamefont {Hassinger}}, \bibinfo {author} {\bibfnamefont {C.}~\bibnamefont {Geibel}}, \ and\ \bibinfo {author} {\bibfnamefont {S.}~\bibnamefont {Khim}},\ }\href {\doibase 10.1103/PhysRevLett.130.166001} {\bibfield  {journal} {\bibinfo  {journal} {Phys. Rev. Lett.}\ }\textbf {\bibinfo {volume} {130}},\ \bibinfo {pages} {166001} (\bibinfo {year} {2023}{\natexlab{b}})}\BibitemShut {NoStop}%
    \bibitem [{\citenamefont {Schertenleib}\ \emph {et~al.}(2021)\citenamefont {Schertenleib}, \citenamefont {Fischer},\ and\ \citenamefont {Sigrist}}]{Schertenleib2021Unusual}%
      \BibitemOpen
      \bibfield  {author} {\bibinfo {author} {\bibfnamefont {E.~G.}\ \bibnamefont {Schertenleib}}, \bibinfo {author} {\bibfnamefont {M.~H.}\ \bibnamefont {Fischer}}, \ and\ \bibinfo {author} {\bibfnamefont {M.}~\bibnamefont {Sigrist}},\ }\href {\doibase 10.1103/PhysRevResearch.3.023179} {\bibfield  {journal} {\bibinfo  {journal} {Phys. Rev. Research}\ }\textbf {\bibinfo {volume} {3}},\ \bibinfo {pages} {023179} (\bibinfo {year} {2021})}\BibitemShut {NoStop}%
    \bibitem [{\citenamefont {Skurativska}\ \emph {et~al.}(2021)\citenamefont {Skurativska}, \citenamefont {Sigrist},\ and\ \citenamefont {Fischer}}]{Skurativska2021Spin}%
      \BibitemOpen
      \bibfield  {author} {\bibinfo {author} {\bibfnamefont {A.}~\bibnamefont {Skurativska}}, \bibinfo {author} {\bibfnamefont {M.}~\bibnamefont {Sigrist}}, \ and\ \bibinfo {author} {\bibfnamefont {M.~H.}\ \bibnamefont {Fischer}},\ }\href {\doibase 10.1103/PhysRevResearch.3.033133} {\bibfield  {journal} {\bibinfo  {journal} {Phys. Rev. Research}\ }\textbf {\bibinfo {volume} {3}},\ \bibinfo {pages} {033133} (\bibinfo {year} {2021})}\BibitemShut {NoStop}%
    \bibitem [{\citenamefont {Nogaki}\ \emph {et~al.}(2021)\citenamefont {Nogaki}, \citenamefont {Daido}, \citenamefont {Ishizuka},\ and\ \citenamefont {Yanase}}]{Nogaki2021topological}%
      \BibitemOpen
      \bibfield  {author} {\bibinfo {author} {\bibfnamefont {K.}~\bibnamefont {Nogaki}}, \bibinfo {author} {\bibfnamefont {A.}~\bibnamefont {Daido}}, \bibinfo {author} {\bibfnamefont {J.}~\bibnamefont {Ishizuka}}, \ and\ \bibinfo {author} {\bibfnamefont {Y.}~\bibnamefont {Yanase}},\ }\href {\doibase 10.1103/PhysRevResearch.3.L032071} {\bibfield  {journal} {\bibinfo  {journal} {Phys. Rev. Research}\ }\textbf {\bibinfo {volume} {3}},\ \bibinfo {pages} {L032071} (\bibinfo {year} {2021})}\BibitemShut {NoStop}%
    \bibitem [{\citenamefont {M\"ockli}\ and\ \citenamefont {Ramires}(2021)}]{Mockli2021Superconductivity}%
      \BibitemOpen
      \bibfield  {author} {\bibinfo {author} {\bibfnamefont {D.}~\bibnamefont {M\"ockli}}\ and\ \bibinfo {author} {\bibfnamefont {A.}~\bibnamefont {Ramires}},\ }\href {\doibase 10.1103/PhysRevB.104.134517} {\bibfield  {journal} {\bibinfo  {journal} {Phys. Rev. B}\ }\textbf {\bibinfo {volume} {104}},\ \bibinfo {pages} {134517} (\bibinfo {year} {2021})}\BibitemShut {NoStop}%
    \bibitem [{\citenamefont {Ptok}\ \emph {et~al.}(2021)\citenamefont {Ptok}, \citenamefont {Kapcia}, \citenamefont {Jochym}, \citenamefont {\L{}a\ifmmode~\dot{z}\else \.{z}\fi{}ewski}, \citenamefont {Ole\ifmmode~\acute{s}\else \'{s}\fi{}},\ and\ \citenamefont {Piekarz}}]{Ptok2021Electronic}%
      \BibitemOpen
      \bibfield  {author} {\bibinfo {author} {\bibfnamefont {A.}~\bibnamefont {Ptok}}, \bibinfo {author} {\bibfnamefont {K.~J.}\ \bibnamefont {Kapcia}}, \bibinfo {author} {\bibfnamefont {P.~T.}\ \bibnamefont {Jochym}}, \bibinfo {author} {\bibfnamefont {J.}~\bibnamefont {\L{}a\ifmmode~\dot{z}\else \.{z}\fi{}ewski}}, \bibinfo {author} {\bibfnamefont {A.~M.}\ \bibnamefont {Ole\ifmmode~\acute{s}\else \'{s}\fi{}}}, \ and\ \bibinfo {author} {\bibfnamefont {P.}~\bibnamefont {Piekarz}},\ }\href {\doibase 10.1103/PhysRevB.104.L041109} {\bibfield  {journal} {\bibinfo  {journal} {Phys. Rev. B}\ }\textbf {\bibinfo {volume} {104}},\ \bibinfo {pages} {L041109} (\bibinfo {year} {2021})}\BibitemShut {NoStop}%
    \bibitem [{\citenamefont {Cavanagh}\ \emph {et~al.}(2022)\citenamefont {Cavanagh}, \citenamefont {Shishidou}, \citenamefont {Weinert}, \citenamefont {Brydon},\ and\ \citenamefont {Agterberg}}]{Cavanagh2022Nonsymmorphic}%
      \BibitemOpen
      \bibfield  {author} {\bibinfo {author} {\bibfnamefont {D.~C.}\ \bibnamefont {Cavanagh}}, \bibinfo {author} {\bibfnamefont {T.}~\bibnamefont {Shishidou}}, \bibinfo {author} {\bibfnamefont {M.}~\bibnamefont {Weinert}}, \bibinfo {author} {\bibfnamefont {P.~M.~R.}\ \bibnamefont {Brydon}}, \ and\ \bibinfo {author} {\bibfnamefont {D.~F.}\ \bibnamefont {Agterberg}},\ }\href {\doibase 10.1103/PhysRevB.105.L020505} {\bibfield  {journal} {\bibinfo  {journal} {Phys. Rev. B}\ }\textbf {\bibinfo {volume} {105}},\ \bibinfo {pages} {L020505} (\bibinfo {year} {2022})}\BibitemShut {NoStop}%
    \bibitem [{\citenamefont {Nogaki}\ and\ \citenamefont {Yanase}(2022)}]{Nogaki2022Even-odd}%
      \BibitemOpen
      \bibfield  {author} {\bibinfo {author} {\bibfnamefont {K.}~\bibnamefont {Nogaki}}\ and\ \bibinfo {author} {\bibfnamefont {Y.}~\bibnamefont {Yanase}},\ }\href {\doibase 10.1103/PhysRevB.106.L100504} {\bibfield  {journal} {\bibinfo  {journal} {Phys. Rev. B}\ }\textbf {\bibinfo {volume} {106}},\ \bibinfo {pages} {L100504} (\bibinfo {year} {2022})}\BibitemShut {NoStop}%
    \bibitem [{\citenamefont {Hazra}\ and\ \citenamefont {Coleman}(2023)}]{Hazra2022Triplet}%
      \BibitemOpen
      \bibfield  {author} {\bibinfo {author} {\bibfnamefont {T.}~\bibnamefont {Hazra}}\ and\ \bibinfo {author} {\bibfnamefont {P.}~\bibnamefont {Coleman}},\ }\href {\doibase 10.1103/PhysRevLett.130.136002} {\bibfield  {journal} {\bibinfo  {journal} {Phys. Rev. Lett.}\ }\textbf {\bibinfo {volume} {130}},\ \bibinfo {pages} {136002} (\bibinfo {year} {2023})}\BibitemShut {NoStop}%
    \bibitem [{\citenamefont {Machida}(2022)}]{Machida2022Violation}%
      \BibitemOpen
      \bibfield  {author} {\bibinfo {author} {\bibfnamefont {K.}~\bibnamefont {Machida}},\ }\href {\doibase 10.1103/PhysRevB.106.184509} {\bibfield  {journal} {\bibinfo  {journal} {Phys. Rev. B}\ }\textbf {\bibinfo {volume} {106}},\ \bibinfo {pages} {184509} (\bibinfo {year} {2022})}\BibitemShut {NoStop}%
    \bibitem [{\citenamefont {Onishi}\ \emph {et~al.}(2022)\citenamefont {Onishi}, \citenamefont {Stockert}, \citenamefont {Khim}, \citenamefont {Banda}, \citenamefont {Brando},\ and\ \citenamefont {Hassinger}}]{Onishi2022Low-Temperature}%
      \BibitemOpen
      \bibfield  {author} {\bibinfo {author} {\bibfnamefont {S.}~\bibnamefont {Onishi}}, \bibinfo {author} {\bibfnamefont {U.}~\bibnamefont {Stockert}}, \bibinfo {author} {\bibfnamefont {S.}~\bibnamefont {Khim}}, \bibinfo {author} {\bibfnamefont {J.}~\bibnamefont {Banda}}, \bibinfo {author} {\bibfnamefont {M.}~\bibnamefont {Brando}}, \ and\ \bibinfo {author} {\bibfnamefont {E.}~\bibnamefont {Hassinger}},\ }\href {\doibase 10.3389/femat.2022.880579} {\bibfield  {journal} {\bibinfo  {journal} {Frontiers in Electronic Materials}\ }\textbf {\bibinfo {volume} {2}} (\bibinfo {year} {2022}),\ 10.3389/femat.2022.880579}\BibitemShut {NoStop}%
    \bibitem [{\citenamefont {Amin}\ \emph {et~al.}(2023)\citenamefont {Amin}, \citenamefont {Wu}, \citenamefont {Shishidou},\ and\ \citenamefont {Agterberg}}]{Amin2023Kramers}%
      \BibitemOpen
      \bibfield  {author} {\bibinfo {author} {\bibfnamefont {A.}~\bibnamefont {Amin}}, \bibinfo {author} {\bibfnamefont {H.}~\bibnamefont {Wu}}, \bibinfo {author} {\bibfnamefont {T.}~\bibnamefont {Shishidou}}, \ and\ \bibinfo {author} {\bibfnamefont {D.~F.}\ \bibnamefont {Agterberg}},\ }\href@noop {} {\enquote {\bibinfo {title} {Kramers' degenerate magnetism and superconductivity},}\ } (\bibinfo {year} {2023}),\ \Eprint {http://arxiv.org/abs/2306.11218} {arXiv:2306.11218 [cond-mat.supr-con]} \BibitemShut {NoStop}%
    \bibitem [{\citenamefont {Sumita}\ and\ \citenamefont {Yanase}(2016)}]{sumita2016}%
      \BibitemOpen
      \bibfield  {author} {\bibinfo {author} {\bibfnamefont {S.}~\bibnamefont {Sumita}}\ and\ \bibinfo {author} {\bibfnamefont {Y.}~\bibnamefont {Yanase}},\ }\href {\doibase 10.1103/PhysRevB.93.224507} {\bibfield  {journal} {\bibinfo  {journal} {Phys. Rev. B}\ }\textbf {\bibinfo {volume} {93}},\ \bibinfo {pages} {224507} (\bibinfo {year} {2016})}\BibitemShut {NoStop}%
    \bibitem [{\citenamefont {Kimura}\ \emph {et~al.}(2021)\citenamefont {Kimura}, \citenamefont {Sichelschmidt},\ and\ \citenamefont {Khim}}]{Kimura2021Optical}%
      \BibitemOpen
      \bibfield  {author} {\bibinfo {author} {\bibfnamefont {S.-i.}\ \bibnamefont {Kimura}}, \bibinfo {author} {\bibfnamefont {J.}~\bibnamefont {Sichelschmidt}}, \ and\ \bibinfo {author} {\bibfnamefont {S.}~\bibnamefont {Khim}},\ }\href {\doibase 10.1103/PhysRevB.104.245116} {\bibfield  {journal} {\bibinfo  {journal} {Phys. Rev. B}\ }\textbf {\bibinfo {volume} {104}},\ \bibinfo {pages} {245116} (\bibinfo {year} {2021})}\BibitemShut {NoStop}%
    \bibitem [{\citenamefont {Qi}\ and\ \citenamefont {Zhang}(2011)}]{Qi2011}%
      \BibitemOpen
      \bibfield  {author} {\bibinfo {author} {\bibfnamefont {X.-L.}\ \bibnamefont {Qi}}\ and\ \bibinfo {author} {\bibfnamefont {S.-C.}\ \bibnamefont {Zhang}},\ }\href {\doibase 10.1103/RevModPhys.83.1057} {\bibfield  {journal} {\bibinfo  {journal} {Rev. Mod. Phys.}\ }\textbf {\bibinfo {volume} {83}},\ \bibinfo {pages} {1057} (\bibinfo {year} {2011})}\BibitemShut {NoStop}%
    \bibitem [{\citenamefont {Sato}\ and\ \citenamefont {Fujimoto}(2016)}]{Sato2016}%
      \BibitemOpen
      \bibfield  {author} {\bibinfo {author} {\bibfnamefont {M.}~\bibnamefont {Sato}}\ and\ \bibinfo {author} {\bibfnamefont {S.}~\bibnamefont {Fujimoto}},\ }\href {\doibase 10.7566/JPSJ.85.072001} {\bibfield  {journal} {\bibinfo  {journal} {J. Phys. Soc. Jpn.}\ }\textbf {\bibinfo {volume} {85}},\ \bibinfo {pages} {072001} (\bibinfo {year} {2016})}\BibitemShut {NoStop}%
    \bibitem [{\citenamefont {Sato}\ and\ \citenamefont {Ando}(2017)}]{Sato2017}%
      \BibitemOpen
      \bibfield  {author} {\bibinfo {author} {\bibfnamefont {M.}~\bibnamefont {Sato}}\ and\ \bibinfo {author} {\bibfnamefont {Y.}~\bibnamefont {Ando}},\ }\href {\doibase 10.1088/1361-6633/aa6ac7} {\bibfield  {journal} {\bibinfo  {journal} {Reports on Progress in Physics}\ }\textbf {\bibinfo {volume} {80}},\ \bibinfo {pages} {076501} (\bibinfo {year} {2017})}\BibitemShut {NoStop}%
    \bibitem [{\citenamefont {Schnyder}\ \emph {et~al.}(2008)\citenamefont {Schnyder}, \citenamefont {Ryu}, \citenamefont {Furusaki},\ and\ \citenamefont {Ludwig}}]{Schnyder2008}%
      \BibitemOpen
      \bibfield  {author} {\bibinfo {author} {\bibfnamefont {A.~P.}\ \bibnamefont {Schnyder}}, \bibinfo {author} {\bibfnamefont {S.}~\bibnamefont {Ryu}}, \bibinfo {author} {\bibfnamefont {A.}~\bibnamefont {Furusaki}}, \ and\ \bibinfo {author} {\bibfnamefont {A.~W.~W.}\ \bibnamefont {Ludwig}},\ }\href {\doibase 10.1103/PhysRevB.78.195125} {\bibfield  {journal} {\bibinfo  {journal} {Phys. Rev. B}\ }\textbf {\bibinfo {volume} {78}},\ \bibinfo {pages} {195125} (\bibinfo {year} {2008})}\BibitemShut {NoStop}%
    \bibitem [{\citenamefont {Kitaev}(2009)}]{Kitaev2009}%
      \BibitemOpen
      \bibfield  {author} {\bibinfo {author} {\bibfnamefont {A.}~\bibnamefont {Kitaev}},\ }\href {\doibase 10.1063/1.3149495} {\bibfield  {journal} {\bibinfo  {journal} {AIP Conf. Proc.}\ }\textbf {\bibinfo {volume} {1134}},\ \bibinfo {pages} {22} (\bibinfo {year} {2009})}\BibitemShut {NoStop}%
    \bibitem [{\citenamefont {Ryu}\ \emph {et~al.}(2010)\citenamefont {Ryu}, \citenamefont {Schnyder}, \citenamefont {Furusaki},\ and\ \citenamefont {Ludwig}}]{Ryu2010}%
      \BibitemOpen
      \bibfield  {author} {\bibinfo {author} {\bibfnamefont {S.}~\bibnamefont {Ryu}}, \bibinfo {author} {\bibfnamefont {A.~P.}\ \bibnamefont {Schnyder}}, \bibinfo {author} {\bibfnamefont {A.}~\bibnamefont {Furusaki}}, \ and\ \bibinfo {author} {\bibfnamefont {A.~W.~W.}\ \bibnamefont {Ludwig}},\ }\href {http://stacks.iop.org/1367-2630/12/i=6/a=065010} {\bibfield  {journal} {\bibinfo  {journal} {New J. Phys.}\ }\textbf {\bibinfo {volume} {12}},\ \bibinfo {pages} {065010} (\bibinfo {year} {2010})}\BibitemShut {NoStop}%
    \bibitem [{\citenamefont {Fu}(2011)}]{Fu2011}%
      \BibitemOpen
      \bibfield  {author} {\bibinfo {author} {\bibfnamefont {L.}~\bibnamefont {Fu}},\ }\href {\doibase 10.1103/PhysRevLett.106.106802} {\bibfield  {journal} {\bibinfo  {journal} {Phys. Rev. Lett.}\ }\textbf {\bibinfo {volume} {106}},\ \bibinfo {pages} {106802} (\bibinfo {year} {2011})}\BibitemShut {NoStop}%
    \bibitem [{\citenamefont {Zhang}\ \emph {et~al.}(2013)\citenamefont {Zhang}, \citenamefont {Kane},\ and\ \citenamefont {Mele}}]{Zhang2013}%
      \BibitemOpen
      \bibfield  {author} {\bibinfo {author} {\bibfnamefont {F.}~\bibnamefont {Zhang}}, \bibinfo {author} {\bibfnamefont {C.~L.}\ \bibnamefont {Kane}}, \ and\ \bibinfo {author} {\bibfnamefont {E.~J.}\ \bibnamefont {Mele}},\ }\href {\doibase 10.1103/PhysRevLett.111.056403} {\bibfield  {journal} {\bibinfo  {journal} {Phys. Rev. Lett.}\ }\textbf {\bibinfo {volume} {111}},\ \bibinfo {pages} {056403} (\bibinfo {year} {2013})}\BibitemShut {NoStop}%
    \bibitem [{\citenamefont {Chiu}\ \emph {et~al.}(2013)\citenamefont {Chiu}, \citenamefont {Yao},\ and\ \citenamefont {Ryu}}]{Chiu2013}%
      \BibitemOpen
      \bibfield  {author} {\bibinfo {author} {\bibfnamefont {C.-K.}\ \bibnamefont {Chiu}}, \bibinfo {author} {\bibfnamefont {H.}~\bibnamefont {Yao}}, \ and\ \bibinfo {author} {\bibfnamefont {S.}~\bibnamefont {Ryu}},\ }\href {\doibase 10.1103/PhysRevB.88.075142} {\bibfield  {journal} {\bibinfo  {journal} {Phys. Rev. B}\ }\textbf {\bibinfo {volume} {88}},\ \bibinfo {pages} {075142} (\bibinfo {year} {2013})}\BibitemShut {NoStop}%
    \bibitem [{\citenamefont {Morimoto}\ and\ \citenamefont {Furusaki}(2013)}]{Morimoto2013}%
      \BibitemOpen
      \bibfield  {author} {\bibinfo {author} {\bibfnamefont {T.}~\bibnamefont {Morimoto}}\ and\ \bibinfo {author} {\bibfnamefont {A.}~\bibnamefont {Furusaki}},\ }\href {\doibase 10.1103/PhysRevB.88.125129} {\bibfield  {journal} {\bibinfo  {journal} {Phys. Rev. B}\ }\textbf {\bibinfo {volume} {88}},\ \bibinfo {pages} {125129} (\bibinfo {year} {2013})}\BibitemShut {NoStop}%
    \bibitem [{\citenamefont {Shiozaki}\ and\ \citenamefont {Sato}(2014)}]{Shiozaki2014}%
      \BibitemOpen
      \bibfield  {author} {\bibinfo {author} {\bibfnamefont {K.}~\bibnamefont {Shiozaki}}\ and\ \bibinfo {author} {\bibfnamefont {M.}~\bibnamefont {Sato}},\ }\href {\doibase 10.1103/PhysRevB.90.165114} {\bibfield  {journal} {\bibinfo  {journal} {Phys. Rev. B}\ }\textbf {\bibinfo {volume} {90}},\ \bibinfo {pages} {165114} (\bibinfo {year} {2014})}\BibitemShut {NoStop}%
    \bibitem [{\citenamefont {Chiu}\ and\ \citenamefont {Schnyder}(2014)}]{Chiu2014}%
      \BibitemOpen
      \bibfield  {author} {\bibinfo {author} {\bibfnamefont {C.-K.}\ \bibnamefont {Chiu}}\ and\ \bibinfo {author} {\bibfnamefont {A.~P.}\ \bibnamefont {Schnyder}},\ }\href {\doibase 10.1103/PhysRevB.90.205136} {\bibfield  {journal} {\bibinfo  {journal} {Phys. Rev. B}\ }\textbf {\bibinfo {volume} {90}},\ \bibinfo {pages} {205136} (\bibinfo {year} {2014})}\BibitemShut {NoStop}%
    \bibitem [{\citenamefont {Ueno}\ \emph {et~al.}(2013)\citenamefont {Ueno}, \citenamefont {Yamakage}, \citenamefont {Tanaka},\ and\ \citenamefont {Sato}}]{Ueno2013}%
      \BibitemOpen
      \bibfield  {author} {\bibinfo {author} {\bibfnamefont {Y.}~\bibnamefont {Ueno}}, \bibinfo {author} {\bibfnamefont {A.}~\bibnamefont {Yamakage}}, \bibinfo {author} {\bibfnamefont {Y.}~\bibnamefont {Tanaka}}, \ and\ \bibinfo {author} {\bibfnamefont {M.}~\bibnamefont {Sato}},\ }\href {\doibase 10.1103/PhysRevLett.111.087002} {\bibfield  {journal} {\bibinfo  {journal} {Phys. Rev. Lett.}\ }\textbf {\bibinfo {volume} {111}},\ \bibinfo {pages} {087002} (\bibinfo {year} {2013})}\BibitemShut {NoStop}%
    \bibitem [{\citenamefont {Tsutsumi}\ \emph {et~al.}(2013)\citenamefont {Tsutsumi}, \citenamefont {Ishikawa}, \citenamefont {Kawakami}, \citenamefont {Mizushima}, \citenamefont {Sato}, \citenamefont {Ichioka},\ and\ \citenamefont {Machida}}]{Tsutsumi2013}%
      \BibitemOpen
      \bibfield  {author} {\bibinfo {author} {\bibfnamefont {Y.}~\bibnamefont {Tsutsumi}}, \bibinfo {author} {\bibfnamefont {M.}~\bibnamefont {Ishikawa}}, \bibinfo {author} {\bibfnamefont {T.}~\bibnamefont {Kawakami}}, \bibinfo {author} {\bibfnamefont {T.}~\bibnamefont {Mizushima}}, \bibinfo {author} {\bibfnamefont {M.}~\bibnamefont {Sato}}, \bibinfo {author} {\bibfnamefont {M.}~\bibnamefont {Ichioka}}, \ and\ \bibinfo {author} {\bibfnamefont {K.}~\bibnamefont {Machida}},\ }\href {\doibase 10.7566/JPSJ.82.113707} {\bibfield  {journal} {\bibinfo  {journal} {J. Phys. Soc. Jpn.}\ }\textbf {\bibinfo {volume} {82}},\ \bibinfo {pages} {113707} (\bibinfo {year} {2013})}\BibitemShut {NoStop}%
    \bibitem [{\citenamefont {Yoshida}\ \emph {et~al.}(2015)\citenamefont {Yoshida}, \citenamefont {Sigrist},\ and\ \citenamefont {Yanase}}]{Yoshida2015}%
      \BibitemOpen
      \bibfield  {author} {\bibinfo {author} {\bibfnamefont {T.}~\bibnamefont {Yoshida}}, \bibinfo {author} {\bibfnamefont {M.}~\bibnamefont {Sigrist}}, \ and\ \bibinfo {author} {\bibfnamefont {Y.}~\bibnamefont {Yanase}},\ }\href {\doibase 10.1103/PhysRevLett.115.027001} {\bibfield  {journal} {\bibinfo  {journal} {Phys. Rev. Lett.}\ }\textbf {\bibinfo {volume} {115}},\ \bibinfo {pages} {027001} (\bibinfo {year} {2015})}\BibitemShut {NoStop}%
    \bibitem [{\citenamefont {Fang}\ and\ \citenamefont {Fu}(2015)}]{Fang2015}%
      \BibitemOpen
      \bibfield  {author} {\bibinfo {author} {\bibfnamefont {C.}~\bibnamefont {Fang}}\ and\ \bibinfo {author} {\bibfnamefont {L.}~\bibnamefont {Fu}},\ }\href {\doibase 10.1103/PhysRevB.91.161105} {\bibfield  {journal} {\bibinfo  {journal} {Phys. Rev. B}\ }\textbf {\bibinfo {volume} {91}},\ \bibinfo {pages} {161105(R)} (\bibinfo {year} {2015})}\BibitemShut {NoStop}%
    \bibitem [{\citenamefont {Shiozaki}\ \emph {et~al.}(2015)\citenamefont {Shiozaki}, \citenamefont {Sato},\ and\ \citenamefont {Gomi}}]{Shiozaki2015}%
      \BibitemOpen
      \bibfield  {author} {\bibinfo {author} {\bibfnamefont {K.}~\bibnamefont {Shiozaki}}, \bibinfo {author} {\bibfnamefont {M.}~\bibnamefont {Sato}}, \ and\ \bibinfo {author} {\bibfnamefont {K.}~\bibnamefont {Gomi}},\ }\href {\doibase 10.1103/PhysRevB.91.155120} {\bibfield  {journal} {\bibinfo  {journal} {Phys. Rev. B}\ }\textbf {\bibinfo {volume} {91}},\ \bibinfo {pages} {155120} (\bibinfo {year} {2015})}\BibitemShut {NoStop}%
    \bibitem [{\citenamefont {Shiozaki}\ \emph {et~al.}(2016)\citenamefont {Shiozaki}, \citenamefont {Sato},\ and\ \citenamefont {Gomi}}]{Shiozaki2016}%
      \BibitemOpen
      \bibfield  {author} {\bibinfo {author} {\bibfnamefont {K.}~\bibnamefont {Shiozaki}}, \bibinfo {author} {\bibfnamefont {M.}~\bibnamefont {Sato}}, \ and\ \bibinfo {author} {\bibfnamefont {K.}~\bibnamefont {Gomi}},\ }\href {\doibase 10.1103/PhysRevB.93.195413} {\bibfield  {journal} {\bibinfo  {journal} {Phys. Rev. B}\ }\textbf {\bibinfo {volume} {93}},\ \bibinfo {pages} {195413} (\bibinfo {year} {2016})}\BibitemShut {NoStop}%
    \bibitem [{\citenamefont {Shapourian}\ \emph {et~al.}(2018)\citenamefont {Shapourian}, \citenamefont {Wang},\ and\ \citenamefont {Ryu}}]{Shapourian2018}%
      \BibitemOpen
      \bibfield  {author} {\bibinfo {author} {\bibfnamefont {H.}~\bibnamefont {Shapourian}}, \bibinfo {author} {\bibfnamefont {Y.}~\bibnamefont {Wang}}, \ and\ \bibinfo {author} {\bibfnamefont {S.}~\bibnamefont {Ryu}},\ }\href {\doibase 10.1103/PhysRevB.97.094508} {\bibfield  {journal} {\bibinfo  {journal} {Phys. Rev. B}\ }\textbf {\bibinfo {volume} {97}},\ \bibinfo {pages} {094508} (\bibinfo {year} {2018})}\BibitemShut {NoStop}%
    \bibitem [{\citenamefont {Yanase}\ and\ \citenamefont {Shiozaki}(2017)}]{Yanase2017}%
      \BibitemOpen
      \bibfield  {author} {\bibinfo {author} {\bibfnamefont {Y.}~\bibnamefont {Yanase}}\ and\ \bibinfo {author} {\bibfnamefont {K.}~\bibnamefont {Shiozaki}},\ }\href {\doibase 10.1103/PhysRevB.95.224514} {\bibfield  {journal} {\bibinfo  {journal} {Phys. Rev. B}\ }\textbf {\bibinfo {volume} {95}},\ \bibinfo {pages} {224514} (\bibinfo {year} {2017})}\BibitemShut {NoStop}%
    \bibitem [{\citenamefont {Daido}\ \emph {et~al.}(2019)\citenamefont {Daido}, \citenamefont {Yoshida},\ and\ \citenamefont {Yanase}}]{Daido2019}%
      \BibitemOpen
      \bibfield  {author} {\bibinfo {author} {\bibfnamefont {A.}~\bibnamefont {Daido}}, \bibinfo {author} {\bibfnamefont {T.}~\bibnamefont {Yoshida}}, \ and\ \bibinfo {author} {\bibfnamefont {Y.}~\bibnamefont {Yanase}},\ }\href {\doibase 10.1103/PhysRevLett.122.227001} {\bibfield  {journal} {\bibinfo  {journal} {Phys. Rev. Lett.}\ }\textbf {\bibinfo {volume} {122}},\ \bibinfo {pages} {227001} (\bibinfo {year} {2019})}\BibitemShut {NoStop}%
    \bibitem [{\citenamefont {Ono}\ \emph {et~al.}(2019)\citenamefont {Ono}, \citenamefont {Yanase},\ and\ \citenamefont {Watanabe}}]{Ono2019}%
      \BibitemOpen
      \bibfield  {author} {\bibinfo {author} {\bibfnamefont {S.}~\bibnamefont {Ono}}, \bibinfo {author} {\bibfnamefont {Y.}~\bibnamefont {Yanase}}, \ and\ \bibinfo {author} {\bibfnamefont {H.}~\bibnamefont {Watanabe}},\ }\href {\doibase 10.1103/PhysRevResearch.1.013012} {\bibfield  {journal} {\bibinfo  {journal} {Phys. Rev. Research}\ }\textbf {\bibinfo {volume} {1}},\ \bibinfo {pages} {013012} (\bibinfo {year} {2019})}\BibitemShut {NoStop}%
    \bibitem [{\citenamefont {Ono}\ \emph {et~al.}(2020)\citenamefont {Ono}, \citenamefont {Po},\ and\ \citenamefont {Watanabe}}]{Ono2020}%
      \BibitemOpen
      \bibfield  {author} {\bibinfo {author} {\bibfnamefont {S.}~\bibnamefont {Ono}}, \bibinfo {author} {\bibfnamefont {H.~C.}\ \bibnamefont {Po}}, \ and\ \bibinfo {author} {\bibfnamefont {H.}~\bibnamefont {Watanabe}},\ }\href {\doibase 10.1126/sciadv.aaz8367} {\bibfield  {journal} {\bibinfo  {journal} {Science Advances}\ }\textbf {\bibinfo {volume} {6}},\ \bibinfo {pages} {eaaz8367} (\bibinfo {year} {2020})}\BibitemShut {NoStop}%
    \bibitem [{\citenamefont {Ono}\ \emph {et~al.}(2021)\citenamefont {Ono}, \citenamefont {Po},\ and\ \citenamefont {Shiozaki}}]{Ono2020_2}%
      \BibitemOpen
      \bibfield  {author} {\bibinfo {author} {\bibfnamefont {S.}~\bibnamefont {Ono}}, \bibinfo {author} {\bibfnamefont {H.~C.}\ \bibnamefont {Po}}, \ and\ \bibinfo {author} {\bibfnamefont {K.}~\bibnamefont {Shiozaki}},\ }\href {\doibase 10.1103/PhysRevResearch.3.023086} {\bibfield  {journal} {\bibinfo  {journal} {Phys. Rev. Research}\ }\textbf {\bibinfo {volume} {3}},\ \bibinfo {pages} {023086(R)} (\bibinfo {year} {2021})}\BibitemShut {NoStop}%
    \bibitem [{\citenamefont {Skurativska}\ \emph {et~al.}(2020)\citenamefont {Skurativska}, \citenamefont {Neupert},\ and\ \citenamefont {Fischer}}]{Skurativska2020}%
      \BibitemOpen
      \bibfield  {author} {\bibinfo {author} {\bibfnamefont {A.}~\bibnamefont {Skurativska}}, \bibinfo {author} {\bibfnamefont {T.}~\bibnamefont {Neupert}}, \ and\ \bibinfo {author} {\bibfnamefont {M.~H.}\ \bibnamefont {Fischer}},\ }\href {\doibase 10.1103/PhysRevResearch.2.013064} {\bibfield  {journal} {\bibinfo  {journal} {Phys. Rev. Research}\ }\textbf {\bibinfo {volume} {2}},\ \bibinfo {pages} {013064} (\bibinfo {year} {2020})}\BibitemShut {NoStop}%
    \bibitem [{\citenamefont {Geier}\ \emph {et~al.}(2020)\citenamefont {Geier}, \citenamefont {Brouwer},\ and\ \citenamefont {Trifunovic}}]{Geier2020}%
      \BibitemOpen
      \bibfield  {author} {\bibinfo {author} {\bibfnamefont {M.}~\bibnamefont {Geier}}, \bibinfo {author} {\bibfnamefont {P.~W.}\ \bibnamefont {Brouwer}}, \ and\ \bibinfo {author} {\bibfnamefont {L.}~\bibnamefont {Trifunovic}},\ }\href {\doibase 10.1103/PhysRevB.101.245128} {\bibfield  {journal} {\bibinfo  {journal} {Phys. Rev. B}\ }\textbf {\bibinfo {volume} {101}},\ \bibinfo {pages} {245128} (\bibinfo {year} {2020})}\BibitemShut {NoStop}%
    \bibitem [{\citenamefont {Shiozaki}(2019)}]{Shiozaki2019}%
      \BibitemOpen
      \bibfield  {author} {\bibinfo {author} {\bibfnamefont {K.}~\bibnamefont {Shiozaki}},\ }\href@noop {} {\enquote {\bibinfo {title} {Variants of the symmetry-based indicator},}\ } (\bibinfo {year} {2019}),\ \Eprint {http://arxiv.org/abs/1907.13632} {arXiv:1907.13632 [cond-mat.mes-hall]} \BibitemShut {NoStop}%
    \bibitem [{\citenamefont {Ahn}\ and\ \citenamefont {Yang}(2020)}]{Ahn2020}%
      \BibitemOpen
      \bibfield  {author} {\bibinfo {author} {\bibfnamefont {J.}~\bibnamefont {Ahn}}\ and\ \bibinfo {author} {\bibfnamefont {B.-J.}\ \bibnamefont {Yang}},\ }\href {\doibase 10.1103/PhysRevResearch.2.012060} {\bibfield  {journal} {\bibinfo  {journal} {Phys. Rev. Research}\ }\textbf {\bibinfo {volume} {2}},\ \bibinfo {pages} {012060(R)} (\bibinfo {year} {2020})}\BibitemShut {NoStop}%
    \bibitem [{\citenamefont {Haule}\ \emph {et~al.}(2010)\citenamefont {Haule}, \citenamefont {Yee},\ and\ \citenamefont {Kim}}]{Haule2010Dynamical}%
      \BibitemOpen
      \bibfield  {author} {\bibinfo {author} {\bibfnamefont {K.}~\bibnamefont {Haule}}, \bibinfo {author} {\bibfnamefont {C.-H.}\ \bibnamefont {Yee}}, \ and\ \bibinfo {author} {\bibfnamefont {K.}~\bibnamefont {Kim}},\ }\href {\doibase 10.1103/PhysRevB.81.195107} {\bibfield  {journal} {\bibinfo  {journal} {Phys. Rev. B}\ }\textbf {\bibinfo {volume} {81}},\ \bibinfo {pages} {195107} (\bibinfo {year} {2010})}\BibitemShut {NoStop}%
    \bibitem [{\citenamefont {Nomoto}\ and\ \citenamefont {Ikeda}(2014)}]{Nomoto2014Fermi}%
      \BibitemOpen
      \bibfield  {author} {\bibinfo {author} {\bibfnamefont {T.}~\bibnamefont {Nomoto}}\ and\ \bibinfo {author} {\bibfnamefont {H.}~\bibnamefont {Ikeda}},\ }\href {\doibase 10.1103/PhysRevB.90.125147} {\bibfield  {journal} {\bibinfo  {journal} {Phys. Rev. B}\ }\textbf {\bibinfo {volume} {90}},\ \bibinfo {pages} {125147} (\bibinfo {year} {2014})}\BibitemShut {NoStop}%
    \bibitem [{\citenamefont {Choi}\ \emph {et~al.}(2012)\citenamefont {Choi}, \citenamefont {Min}, \citenamefont {Shim}, \citenamefont {Haule},\ and\ \citenamefont {Kotliar}}]{Choi2012Temperature}%
      \BibitemOpen
      \bibfield  {author} {\bibinfo {author} {\bibfnamefont {H.~C.}\ \bibnamefont {Choi}}, \bibinfo {author} {\bibfnamefont {B.~I.}\ \bibnamefont {Min}}, \bibinfo {author} {\bibfnamefont {J.~H.}\ \bibnamefont {Shim}}, \bibinfo {author} {\bibfnamefont {K.}~\bibnamefont {Haule}}, \ and\ \bibinfo {author} {\bibfnamefont {G.}~\bibnamefont {Kotliar}},\ }\href {\doibase 10.1103/PhysRevLett.108.016402} {\bibfield  {journal} {\bibinfo  {journal} {Phys. Rev. Lett.}\ }\textbf {\bibinfo {volume} {108}},\ \bibinfo {pages} {016402} (\bibinfo {year} {2012})}\BibitemShut {NoStop}%
    \bibitem [{\citenamefont {Ikeda}\ \emph {et~al.}(2015)\citenamefont {Ikeda}, \citenamefont {Suzuki},\ and\ \citenamefont {Arita}}]{Ikeda2015Emergent}%
      \BibitemOpen
      \bibfield  {author} {\bibinfo {author} {\bibfnamefont {H.}~\bibnamefont {Ikeda}}, \bibinfo {author} {\bibfnamefont {M.-T.}\ \bibnamefont {Suzuki}}, \ and\ \bibinfo {author} {\bibfnamefont {R.}~\bibnamefont {Arita}},\ }\href {\doibase 10.1103/PhysRevLett.114.147003} {\bibfield  {journal} {\bibinfo  {journal} {Phys. Rev. Lett.}\ }\textbf {\bibinfo {volume} {114}},\ \bibinfo {pages} {147003} (\bibinfo {year} {2015})}\BibitemShut {NoStop}%
    \bibitem [{\citenamefont {Ishizuka}\ \emph {et~al.}(2019)\citenamefont {Ishizuka}, \citenamefont {Sumita}, \citenamefont {Daido},\ and\ \citenamefont {Yanase}}]{Ishizuka2019Insulator-Metal}%
      \BibitemOpen
      \bibfield  {author} {\bibinfo {author} {\bibfnamefont {J.}~\bibnamefont {Ishizuka}}, \bibinfo {author} {\bibfnamefont {S.}~\bibnamefont {Sumita}}, \bibinfo {author} {\bibfnamefont {A.}~\bibnamefont {Daido}}, \ and\ \bibinfo {author} {\bibfnamefont {Y.}~\bibnamefont {Yanase}},\ }\href {\doibase 10.1103/PhysRevLett.123.217001} {\bibfield  {journal} {\bibinfo  {journal} {Phys. Rev. Lett.}\ }\textbf {\bibinfo {volume} {123}},\ \bibinfo {pages} {217001} (\bibinfo {year} {2019})}\BibitemShut {NoStop}%
    \bibitem [{\citenamefont {Xu}\ \emph {et~al.}(2019)\citenamefont {Xu}, \citenamefont {Sheng},\ and\ \citenamefont {Yang}}]{Xu2019Quasi-Two-Dimensional}%
      \BibitemOpen
      \bibfield  {author} {\bibinfo {author} {\bibfnamefont {Y.}~\bibnamefont {Xu}}, \bibinfo {author} {\bibfnamefont {Y.}~\bibnamefont {Sheng}}, \ and\ \bibinfo {author} {\bibfnamefont {Y.-f.}\ \bibnamefont {Yang}},\ }\href {\doibase 10.1103/PhysRevLett.123.217002} {\bibfield  {journal} {\bibinfo  {journal} {Phys. Rev. Lett.}\ }\textbf {\bibinfo {volume} {123}},\ \bibinfo {pages} {217002} (\bibinfo {year} {2019})}\BibitemShut {NoStop}%
    \bibitem [{\citenamefont {Chen}\ \emph {et~al.}(2023)\citenamefont {Chen}, \citenamefont {Wang}, \citenamefont {Ishizuka}, \citenamefont {Nogaki}, \citenamefont {Cheng}, \citenamefont {Yang}, \citenamefont {Zhang}, \citenamefont {Chen}, \citenamefont {Zhu}, \citenamefont {Yanase}, \citenamefont {Lv},\ and\ \citenamefont {Huang}}]{Chen2023Coexistence}%
      \BibitemOpen
      \bibfield  {author} {\bibinfo {author} {\bibfnamefont {X.}~\bibnamefont {Chen}}, \bibinfo {author} {\bibfnamefont {L.}~\bibnamefont {Wang}}, \bibinfo {author} {\bibfnamefont {J.}~\bibnamefont {Ishizuka}}, \bibinfo {author} {\bibfnamefont {K.}~\bibnamefont {Nogaki}}, \bibinfo {author} {\bibfnamefont {Y.}~\bibnamefont {Cheng}}, \bibinfo {author} {\bibfnamefont {F.}~\bibnamefont {Yang}}, \bibinfo {author} {\bibfnamefont {R.}~\bibnamefont {Zhang}}, \bibinfo {author} {\bibfnamefont {Z.}~\bibnamefont {Chen}}, \bibinfo {author} {\bibfnamefont {F.}~\bibnamefont {Zhu}}, \bibinfo {author} {\bibfnamefont {Y.}~\bibnamefont {Yanase}}, \bibinfo {author} {\bibfnamefont {B.}~\bibnamefont {Lv}}, \ and\ \bibinfo {author} {\bibfnamefont {Y.}~\bibnamefont {Huang}},\ }\href@noop {} {\enquote {\bibinfo {title} {Coexistence of near-ef flat band and van hove singularity in a two-phase superconductor},}\ } (\bibinfo {year} {2023}),\ \Eprint {http://arxiv.org/abs/2309.05895} {arXiv:2309.05895 [cond-mat.str-el]} \BibitemShut
      {NoStop}%
    \bibitem [{\citenamefont {Wu}\ \emph {et~al.}(2023)\citenamefont {Wu}, \citenamefont {Zhang}, \citenamefont {Ju}, \citenamefont {Hu}, \citenamefont {Yang}, \citenamefont {Zheng}, \citenamefont {Huang}, \citenamefont {Zhang}, \citenamefont {Zhang}, \citenamefont {Song}, \citenamefont {Plumb}, \citenamefont {Steglich}, \citenamefont {Shi}, \citenamefont {Zwicknagl}, \citenamefont {Cao}, \citenamefont {Yuan},\ and\ \citenamefont {Liu}}]{Wu2023Quasi}%
      \BibitemOpen
      \bibfield  {author} {\bibinfo {author} {\bibfnamefont {Y.}~\bibnamefont {Wu}}, \bibinfo {author} {\bibfnamefont {Y.}~\bibnamefont {Zhang}}, \bibinfo {author} {\bibfnamefont {S.}~\bibnamefont {Ju}}, \bibinfo {author} {\bibfnamefont {Y.}~\bibnamefont {Hu}}, \bibinfo {author} {\bibfnamefont {G.}~\bibnamefont {Yang}}, \bibinfo {author} {\bibfnamefont {H.}~\bibnamefont {Zheng}}, \bibinfo {author} {\bibfnamefont {Y.}~\bibnamefont {Huang}}, \bibinfo {author} {\bibfnamefont {Y.}~\bibnamefont {Zhang}}, \bibinfo {author} {\bibfnamefont {H.}~\bibnamefont {Zhang}}, \bibinfo {author} {\bibfnamefont {B.}~\bibnamefont {Song}}, \bibinfo {author} {\bibfnamefont {N.~C.}\ \bibnamefont {Plumb}}, \bibinfo {author} {\bibfnamefont {F.}~\bibnamefont {Steglich}}, \bibinfo {author} {\bibfnamefont {M.}~\bibnamefont {Shi}}, \bibinfo {author} {\bibfnamefont {G.}~\bibnamefont {Zwicknagl}}, \bibinfo {author} {\bibfnamefont {C.}~\bibnamefont {Cao}}, \bibinfo {author} {\bibfnamefont {H.}~\bibnamefont {Yuan}}, \ and\ \bibinfo {author}
      {\bibfnamefont {Y.}~\bibnamefont {Liu}},\ }\href@noop {} {\enquote {\bibinfo {title} {Quasi-two-dimensional fermi surface and heavy quasiparticles in cerh2as2},}\ } (\bibinfo {year} {2023}),\ \Eprint {http://arxiv.org/abs/2309.06732} {arXiv:2309.06732 [cond-mat.str-el]} \BibitemShut {NoStop}%
    \bibitem [{sup()}]{suppl}%
      \BibitemOpen
      \href@noop {} {}\bibinfo {note} {See Supplemental Material for details, which includes Refs.~\cite{blaha_2, Czyzyk_LDA+U_AMF, Wigner, Herring1937, Inui-Tanabe-Onodera, Bradley, Shiozaki2022, Kunes2010, Pizzi2020, Zhi2022}.}\BibitemShut {Stop}%
    \bibitem [{\citenamefont {Mazin}\ \emph {et~al.}(2008)\citenamefont {Mazin}, \citenamefont {Singh}, \citenamefont {Johannes},\ and\ \citenamefont {Du}}]{Mazin2008}%
      \BibitemOpen
      \bibfield  {author} {\bibinfo {author} {\bibfnamefont {I.~I.}\ \bibnamefont {Mazin}}, \bibinfo {author} {\bibfnamefont {D.~J.}\ \bibnamefont {Singh}}, \bibinfo {author} {\bibfnamefont {M.~D.}\ \bibnamefont {Johannes}}, \ and\ \bibinfo {author} {\bibfnamefont {M.~H.}\ \bibnamefont {Du}},\ }\href {\doibase 10.1103/PhysRevLett.101.057003} {\bibfield  {journal} {\bibinfo  {journal} {Phys. Rev. Lett.}\ }\textbf {\bibinfo {volume} {101}},\ \bibinfo {pages} {057003} (\bibinfo {year} {2008})}\BibitemShut {NoStop}%
    \bibitem [{\citenamefont {Kuroki}\ \emph {et~al.}(2008)\citenamefont {Kuroki}, \citenamefont {Onari}, \citenamefont {Arita}, \citenamefont {Usui}, \citenamefont {Tanaka}, \citenamefont {Kontani},\ and\ \citenamefont {Aoki}}]{Kuroki2008}%
      \BibitemOpen
      \bibfield  {author} {\bibinfo {author} {\bibfnamefont {K.}~\bibnamefont {Kuroki}}, \bibinfo {author} {\bibfnamefont {S.}~\bibnamefont {Onari}}, \bibinfo {author} {\bibfnamefont {R.}~\bibnamefont {Arita}}, \bibinfo {author} {\bibfnamefont {H.}~\bibnamefont {Usui}}, \bibinfo {author} {\bibfnamefont {Y.}~\bibnamefont {Tanaka}}, \bibinfo {author} {\bibfnamefont {H.}~\bibnamefont {Kontani}}, \ and\ \bibinfo {author} {\bibfnamefont {H.}~\bibnamefont {Aoki}},\ }\href {\doibase 10.1103/PhysRevLett.101.087004} {\bibfield  {journal} {\bibinfo  {journal} {Phys. Rev. Lett.}\ }\textbf {\bibinfo {volume} {101}},\ \bibinfo {pages} {087004} (\bibinfo {year} {2008})}\BibitemShut {NoStop}%
    \bibitem [{\citenamefont {Fernandes}\ \emph {et~al.}(2010)\citenamefont {Fernandes}, \citenamefont {VanBebber}, \citenamefont {Bhattacharya}, \citenamefont {Chandra}, \citenamefont {Keppens}, \citenamefont {Mandrus}, \citenamefont {McGuire}, \citenamefont {Sales}, \citenamefont {Sefat},\ and\ \citenamefont {Schmalian}}]{Fernandes2010}%
      \BibitemOpen
      \bibfield  {author} {\bibinfo {author} {\bibfnamefont {R.~M.}\ \bibnamefont {Fernandes}}, \bibinfo {author} {\bibfnamefont {L.~H.}\ \bibnamefont {VanBebber}}, \bibinfo {author} {\bibfnamefont {S.}~\bibnamefont {Bhattacharya}}, \bibinfo {author} {\bibfnamefont {P.}~\bibnamefont {Chandra}}, \bibinfo {author} {\bibfnamefont {V.}~\bibnamefont {Keppens}}, \bibinfo {author} {\bibfnamefont {D.}~\bibnamefont {Mandrus}}, \bibinfo {author} {\bibfnamefont {M.~A.}\ \bibnamefont {McGuire}}, \bibinfo {author} {\bibfnamefont {B.~C.}\ \bibnamefont {Sales}}, \bibinfo {author} {\bibfnamefont {A.~S.}\ \bibnamefont {Sefat}}, \ and\ \bibinfo {author} {\bibfnamefont {J.}~\bibnamefont {Schmalian}},\ }\href {\doibase 10.1103/PhysRevLett.105.157003} {\bibfield  {journal} {\bibinfo  {journal} {Phys. Rev. Lett.}\ }\textbf {\bibinfo {volume} {105}},\ \bibinfo {pages} {157003} (\bibinfo {year} {2010})}\BibitemShut {NoStop}%
    \bibitem [{\citenamefont {Yanagi}\ \emph {et~al.}(2010)\citenamefont {Yanagi}, \citenamefont {Yamakawa}, \citenamefont {Adachi},\ and\ \citenamefont {\ifmmode~\bar{O}\else \={O}\fi{}no}}]{Yanagi2010}%
      \BibitemOpen
      \bibfield  {author} {\bibinfo {author} {\bibfnamefont {Y.}~\bibnamefont {Yanagi}}, \bibinfo {author} {\bibfnamefont {Y.}~\bibnamefont {Yamakawa}}, \bibinfo {author} {\bibfnamefont {N.}~\bibnamefont {Adachi}}, \ and\ \bibinfo {author} {\bibfnamefont {Y.}~\bibnamefont {\ifmmode~\bar{O}\else \={O}\fi{}no}},\ }\href {\doibase 10.1143/JPSJ.79.123707} {\bibfield  {journal} {\bibinfo  {journal} {J. Phys. Soc. Jpn.}\ }\textbf {\bibinfo {volume} {79}},\ \bibinfo {pages} {123707} (\bibinfo {year} {2010})}\BibitemShut {NoStop}%
    \bibitem [{\citenamefont {Kontani}\ \emph {et~al.}(2011)\citenamefont {Kontani}, \citenamefont {Saito},\ and\ \citenamefont {Onari}}]{Kontani2011}%
      \BibitemOpen
      \bibfield  {author} {\bibinfo {author} {\bibfnamefont {H.}~\bibnamefont {Kontani}}, \bibinfo {author} {\bibfnamefont {T.}~\bibnamefont {Saito}}, \ and\ \bibinfo {author} {\bibfnamefont {S.}~\bibnamefont {Onari}},\ }\href {\doibase 10.1103/PhysRevB.84.024528} {\bibfield  {journal} {\bibinfo  {journal} {Phys. Rev. B}\ }\textbf {\bibinfo {volume} {84}},\ \bibinfo {pages} {024528} (\bibinfo {year} {2011})}\BibitemShut {NoStop}%
    \bibitem [{\citenamefont {Onari}\ and\ \citenamefont {Kontani}(2012)}]{Onari2012}%
      \BibitemOpen
      \bibfield  {author} {\bibinfo {author} {\bibfnamefont {S.}~\bibnamefont {Onari}}\ and\ \bibinfo {author} {\bibfnamefont {H.}~\bibnamefont {Kontani}},\ }\href {\doibase 10.1103/PhysRevLett.109.137001} {\bibfield  {journal} {\bibinfo  {journal} {Phys. Rev. Lett.}\ }\textbf {\bibinfo {volume} {109}},\ \bibinfo {pages} {137001} (\bibinfo {year} {2012})}\BibitemShut {NoStop}%
    \bibitem [{\citenamefont {Fu}\ and\ \citenamefont {Berg}(2010)}]{Fu2010}%
      \BibitemOpen
      \bibfield  {author} {\bibinfo {author} {\bibfnamefont {L.}~\bibnamefont {Fu}}\ and\ \bibinfo {author} {\bibfnamefont {E.}~\bibnamefont {Berg}},\ }\href {\doibase 10.1103/PhysRevLett.105.097001} {\bibfield  {journal} {\bibinfo  {journal} {Phys. Rev. Lett.}\ }\textbf {\bibinfo {volume} {105}},\ \bibinfo {pages} {097001} (\bibinfo {year} {2010})}\BibitemShut {NoStop}%
    \bibitem [{\citenamefont {Sato}(2010)}]{Sato2010}%
      \BibitemOpen
      \bibfield  {author} {\bibinfo {author} {\bibfnamefont {M.}~\bibnamefont {Sato}},\ }\href {\doibase 10.1103/PhysRevB.81.220504} {\bibfield  {journal} {\bibinfo  {journal} {Phys. Rev. B}\ }\textbf {\bibinfo {volume} {81}},\ \bibinfo {pages} {220504(R)} (\bibinfo {year} {2010})}\BibitemShut {NoStop}%
    \bibitem [{\citenamefont {Yanase}(2016)}]{Yanase2016}%
      \BibitemOpen
      \bibfield  {author} {\bibinfo {author} {\bibfnamefont {Y.}~\bibnamefont {Yanase}},\ }\href {\doibase 10.1103/PhysRevB.94.174502} {\bibfield  {journal} {\bibinfo  {journal} {Phys. Rev. B}\ }\textbf {\bibinfo {volume} {94}},\ \bibinfo {pages} {174502} (\bibinfo {year} {2016})}\BibitemShut {NoStop}%
    \bibitem [{\citenamefont {Sigrist}\ and\ \citenamefont {Ueda}(1991)}]{Sigrist-Ueda}%
      \BibitemOpen
      \bibfield  {author} {\bibinfo {author} {\bibfnamefont {M.}~\bibnamefont {Sigrist}}\ and\ \bibinfo {author} {\bibfnamefont {K.}~\bibnamefont {Ueda}},\ }\href {\doibase 10.1103/RevModPhys.63.239} {\bibfield  {journal} {\bibinfo  {journal} {Rev. Mod. Phys.}\ }\textbf {\bibinfo {volume} {63}},\ \bibinfo {pages} {239} (\bibinfo {year} {1991})}\BibitemShut {NoStop}%
    \bibitem [{\citenamefont {Kobayashi}\ \emph {et~al.}(2014)\citenamefont {Kobayashi}, \citenamefont {Shiozaki}, \citenamefont {Tanaka},\ and\ \citenamefont {Sato}}]{Kobayashi2014}%
      \BibitemOpen
      \bibfield  {author} {\bibinfo {author} {\bibfnamefont {S.}~\bibnamefont {Kobayashi}}, \bibinfo {author} {\bibfnamefont {K.}~\bibnamefont {Shiozaki}}, \bibinfo {author} {\bibfnamefont {Y.}~\bibnamefont {Tanaka}}, \ and\ \bibinfo {author} {\bibfnamefont {M.}~\bibnamefont {Sato}},\ }\href {\doibase 10.1103/PhysRevB.90.024516} {\bibfield  {journal} {\bibinfo  {journal} {Phys. Rev. B}\ }\textbf {\bibinfo {volume} {90}},\ \bibinfo {pages} {024516} (\bibinfo {year} {2014})}\BibitemShut {NoStop}%
    \bibitem [{\citenamefont {Kobayashi}\ \emph {et~al.}(2016)\citenamefont {Kobayashi}, \citenamefont {Yanase},\ and\ \citenamefont {Sato}}]{Kobayashi2016}%
      \BibitemOpen
      \bibfield  {author} {\bibinfo {author} {\bibfnamefont {S.}~\bibnamefont {Kobayashi}}, \bibinfo {author} {\bibfnamefont {Y.}~\bibnamefont {Yanase}}, \ and\ \bibinfo {author} {\bibfnamefont {M.}~\bibnamefont {Sato}},\ }\href {\doibase 10.1103/PhysRevB.94.134512} {\bibfield  {journal} {\bibinfo  {journal} {Phys. Rev. B}\ }\textbf {\bibinfo {volume} {94}},\ \bibinfo {pages} {134512} (\bibinfo {year} {2016})}\BibitemShut {NoStop}%
    \bibitem [{\citenamefont {Kobayashi}\ \emph {et~al.}(2018)\citenamefont {Kobayashi}, \citenamefont {Sumita}, \citenamefont {Yanase},\ and\ \citenamefont {Sato}}]{Kobayashi2018}%
      \BibitemOpen
      \bibfield  {author} {\bibinfo {author} {\bibfnamefont {S.}~\bibnamefont {Kobayashi}}, \bibinfo {author} {\bibfnamefont {S.}~\bibnamefont {Sumita}}, \bibinfo {author} {\bibfnamefont {Y.}~\bibnamefont {Yanase}}, \ and\ \bibinfo {author} {\bibfnamefont {M.}~\bibnamefont {Sato}},\ }\href {\doibase 10.1103/PhysRevB.97.180504} {\bibfield  {journal} {\bibinfo  {journal} {Phys. Rev. B}\ }\textbf {\bibinfo {volume} {97}},\ \bibinfo {pages} {180504(R)} (\bibinfo {year} {2018})}\BibitemShut {NoStop}%
    \bibitem [{\citenamefont {Sumita}\ \emph {et~al.}(2019)\citenamefont {Sumita}, \citenamefont {Nomoto}, \citenamefont {Shiozaki},\ and\ \citenamefont {Yanase}}]{Sumita2019}%
      \BibitemOpen
      \bibfield  {author} {\bibinfo {author} {\bibfnamefont {S.}~\bibnamefont {Sumita}}, \bibinfo {author} {\bibfnamefont {T.}~\bibnamefont {Nomoto}}, \bibinfo {author} {\bibfnamefont {K.}~\bibnamefont {Shiozaki}}, \ and\ \bibinfo {author} {\bibfnamefont {Y.}~\bibnamefont {Yanase}},\ }\href {\doibase 10.1103/PhysRevB.99.134513} {\bibfield  {journal} {\bibinfo  {journal} {Phys. Rev. B}\ }\textbf {\bibinfo {volume} {99}},\ \bibinfo {pages} {134513} (\bibinfo {year} {2019})}\BibitemShut {NoStop}%
    \bibitem [{\citenamefont {Ono}\ and\ \citenamefont {Shiozaki}(2022)}]{Ono2022Symmetry-Based}%
      \BibitemOpen
      \bibfield  {author} {\bibinfo {author} {\bibfnamefont {S.}~\bibnamefont {Ono}}\ and\ \bibinfo {author} {\bibfnamefont {K.}~\bibnamefont {Shiozaki}},\ }\href {\doibase 10.1103/PhysRevX.12.011021} {\bibfield  {journal} {\bibinfo  {journal} {Phys. Rev. X}\ }\textbf {\bibinfo {volume} {12}},\ \bibinfo {pages} {011021} (\bibinfo {year} {2022})}\BibitemShut {NoStop}%
    \bibitem [{\citenamefont {Yarzhemsky}\ and\ \citenamefont {Murav'ev}(1992)}]{Yarzhemsky1992}%
      \BibitemOpen
      \bibfield  {author} {\bibinfo {author} {\bibfnamefont {V.~G.}\ \bibnamefont {Yarzhemsky}}\ and\ \bibinfo {author} {\bibfnamefont {E.~N.}\ \bibnamefont {Murav'ev}},\ }\href {http://stacks.iop.org/0953-8984/4/i=13/a=015} {\bibfield  {journal} {\bibinfo  {journal} {J. Phys.: Condens. Matter}\ }\textbf {\bibinfo {volume} {4}},\ \bibinfo {pages} {3525} (\bibinfo {year} {1992})}\BibitemShut {NoStop}%
    \bibitem [{\citenamefont {Norman}(1995)}]{Norman1995}%
      \BibitemOpen
      \bibfield  {author} {\bibinfo {author} {\bibfnamefont {M.~R.}\ \bibnamefont {Norman}},\ }\href {\doibase 10.1103/PhysRevB.52.15093} {\bibfield  {journal} {\bibinfo  {journal} {Phys. Rev. B}\ }\textbf {\bibinfo {volume} {52}},\ \bibinfo {pages} {15093} (\bibinfo {year} {1995})}\BibitemShut {NoStop}%
    \bibitem [{\citenamefont {Micklitz}\ and\ \citenamefont {Norman}(2009)}]{Micklitz2009}%
      \BibitemOpen
      \bibfield  {author} {\bibinfo {author} {\bibfnamefont {T.}~\bibnamefont {Micklitz}}\ and\ \bibinfo {author} {\bibfnamefont {M.~R.}\ \bibnamefont {Norman}},\ }\href {\doibase 10.1103/PhysRevB.80.100506} {\bibfield  {journal} {\bibinfo  {journal} {Phys. Rev. B}\ }\textbf {\bibinfo {volume} {80}},\ \bibinfo {pages} {100506(R)} (\bibinfo {year} {2009})}\BibitemShut {NoStop}%
    \bibitem [{\citenamefont {Nomoto}\ and\ \citenamefont {Ikeda}(2017)}]{Nomoto2017}%
      \BibitemOpen
      \bibfield  {author} {\bibinfo {author} {\bibfnamefont {T.}~\bibnamefont {Nomoto}}\ and\ \bibinfo {author} {\bibfnamefont {H.}~\bibnamefont {Ikeda}},\ }\href {\doibase 10.7566/JPSJ.86.023703} {\bibfield  {journal} {\bibinfo  {journal} {J. Phys. Soc. Jpn.}\ }\textbf {\bibinfo {volume} {86}},\ \bibinfo {pages} {023703} (\bibinfo {year} {2017})}\BibitemShut {NoStop}%
    \bibitem [{\citenamefont {Micklitz}\ and\ \citenamefont {Norman}(2017)}]{Micklitz2017_PRL}%
      \BibitemOpen
      \bibfield  {author} {\bibinfo {author} {\bibfnamefont {T.}~\bibnamefont {Micklitz}}\ and\ \bibinfo {author} {\bibfnamefont {M.~R.}\ \bibnamefont {Norman}},\ }\href {\doibase 10.1103/PhysRevLett.118.207001} {\bibfield  {journal} {\bibinfo  {journal} {Phys. Rev. Lett.}\ }\textbf {\bibinfo {volume} {118}},\ \bibinfo {pages} {207001} (\bibinfo {year} {2017})}\BibitemShut {NoStop}%
    \bibitem [{\citenamefont {Sumita}\ \emph {et~al.}(2017)\citenamefont {Sumita}, \citenamefont {Nomoto},\ and\ \citenamefont {Yanase}}]{Sumita2017}%
      \BibitemOpen
      \bibfield  {author} {\bibinfo {author} {\bibfnamefont {S.}~\bibnamefont {Sumita}}, \bibinfo {author} {\bibfnamefont {T.}~\bibnamefont {Nomoto}}, \ and\ \bibinfo {author} {\bibfnamefont {Y.}~\bibnamefont {Yanase}},\ }\href {\doibase 10.1103/PhysRevLett.119.027001} {\bibfield  {journal} {\bibinfo  {journal} {Phys. Rev. Lett.}\ }\textbf {\bibinfo {volume} {119}},\ \bibinfo {pages} {027001} (\bibinfo {year} {2017})}\BibitemShut {NoStop}%
    \bibitem [{\citenamefont {Sumita}\ and\ \citenamefont {Yanase}(2018)}]{Sumita2018}%
      \BibitemOpen
      \bibfield  {author} {\bibinfo {author} {\bibfnamefont {S.}~\bibnamefont {Sumita}}\ and\ \bibinfo {author} {\bibfnamefont {Y.}~\bibnamefont {Yanase}},\ }\href {\doibase 10.1103/PhysRevB.97.134512} {\bibfield  {journal} {\bibinfo  {journal} {Phys. Rev. B}\ }\textbf {\bibinfo {volume} {97}},\ \bibinfo {pages} {134512} (\bibinfo {year} {2018})}\BibitemShut {NoStop}%
    \bibitem [{bla()}]{blaha_2}%
      \BibitemOpen
      \href@noop {} {}\bibinfo {note} {P. Blaha, K. Schwarz, G. K. H. Madsen, D. Kvasnicka, J. Luitz, R. Laskowski, F. Tran, and L. D. Marks, \textit{WIEN2k, An Augmented Plane Wave + Local Orbitals Program for Calculating Crystal Properties} (Karlheinz Schwarz, Techn. Universit\"at Wien, Austria, 2018).}\BibitemShut {Stop}%
    \bibitem [{\citenamefont {Czy\ifmmode~\dot{z}\else \.{z}\fi{}yk}\ and\ \citenamefont {Sawatzky}(1994)}]{Czyzyk_LDA+U_AMF}%
      \BibitemOpen
      \bibfield  {author} {\bibinfo {author} {\bibfnamefont {M.~T.}\ \bibnamefont {Czy\ifmmode~\dot{z}\else \.{z}\fi{}yk}}\ and\ \bibinfo {author} {\bibfnamefont {G.~A.}\ \bibnamefont {Sawatzky}},\ }\href {\doibase 10.1103/PhysRevB.49.14211} {\bibfield  {journal} {\bibinfo  {journal} {Phys. Rev. B}\ }\textbf {\bibinfo {volume} {49}},\ \bibinfo {pages} {14211} (\bibinfo {year} {1994})}\BibitemShut {NoStop}%
    \bibitem [{\citenamefont {Wigner}(1959)}]{Wigner}%
      \BibitemOpen
      \bibfield  {author} {\bibinfo {author} {\bibfnamefont {E.~P.}\ \bibnamefont {Wigner}},\ }\href@noop {} {\emph {\bibinfo {title} {Group Theory and Its Application to the Quantum Mechanics of Atomic Spectra}}}\ (\bibinfo  {publisher} {Academic Press},\ \bibinfo {address} {New York},\ \bibinfo {year} {1959})\BibitemShut {NoStop}%
    \bibitem [{\citenamefont {Herring}(1937)}]{Herring1937}%
      \BibitemOpen
      \bibfield  {author} {\bibinfo {author} {\bibfnamefont {C.}~\bibnamefont {Herring}},\ }\href {\doibase 10.1103/PhysRev.52.361} {\bibfield  {journal} {\bibinfo  {journal} {Phys. Rev.}\ }\textbf {\bibinfo {volume} {52}},\ \bibinfo {pages} {361} (\bibinfo {year} {1937})}\BibitemShut {NoStop}%
    \bibitem [{\citenamefont {Inui}\ \emph {et~al.}(1990)\citenamefont {Inui}, \citenamefont {Tanabe},\ and\ \citenamefont {Onodera}}]{Inui-Tanabe-Onodera}%
      \BibitemOpen
      \bibfield  {author} {\bibinfo {author} {\bibfnamefont {T.}~\bibnamefont {Inui}}, \bibinfo {author} {\bibfnamefont {Y.}~\bibnamefont {Tanabe}}, \ and\ \bibinfo {author} {\bibfnamefont {Y.}~\bibnamefont {Onodera}},\ }\href@noop {} {\emph {\bibinfo {title} {Group Theory and Its Applications in Physics}}},\ \bibinfo {series} {Springer Series in Solid-State Sciences}, Vol.~\bibinfo {volume} {78}\ (\bibinfo  {publisher} {Springer-Verlag Berlin Heidelberg},\ \bibinfo {address} {Berlin, Heidelberg},\ \bibinfo {year} {1990})\BibitemShut {NoStop}%
    \bibitem [{\citenamefont {Bradley}\ and\ \citenamefont {Cracknell}(1972)}]{Bradley}%
      \BibitemOpen
      \bibfield  {author} {\bibinfo {author} {\bibfnamefont {C.~J.}\ \bibnamefont {Bradley}}\ and\ \bibinfo {author} {\bibfnamefont {A.~P.}\ \bibnamefont {Cracknell}},\ }\href@noop {} {\emph {\bibinfo {title} {The Mathematical Theory of Symmetry in Solids}}}\ (\bibinfo  {publisher} {Oxford University Press},\ \bibinfo {address} {Oxford},\ \bibinfo {year} {1972})\BibitemShut {NoStop}%
    \bibitem [{\citenamefont {Shiozaki}\ \emph {et~al.}(2022)\citenamefont {Shiozaki}, \citenamefont {Sato},\ and\ \citenamefont {Gomi}}]{Shiozaki2022}%
      \BibitemOpen
      \bibfield  {author} {\bibinfo {author} {\bibfnamefont {K.}~\bibnamefont {Shiozaki}}, \bibinfo {author} {\bibfnamefont {M.}~\bibnamefont {Sato}}, \ and\ \bibinfo {author} {\bibfnamefont {K.}~\bibnamefont {Gomi}},\ }\href {\doibase 10.1103/PhysRevB.106.165103} {\bibfield  {journal} {\bibinfo  {journal} {Phys. Rev. B}\ }\textbf {\bibinfo {volume} {106}},\ \bibinfo {pages} {165103} (\bibinfo {year} {2022})}\BibitemShut {NoStop}%
    \bibitem [{\citenamefont {Kuneš}\ \emph {et~al.}(2010)\citenamefont {Kuneš}, \citenamefont {Arita}, \citenamefont {Wissgott}, \citenamefont {Toschi}, \citenamefont {Ikeda},\ and\ \citenamefont {Held}}]{Kunes2010}%
      \BibitemOpen
      \bibfield  {author} {\bibinfo {author} {\bibfnamefont {J.}~\bibnamefont {Kuneš}}, \bibinfo {author} {\bibfnamefont {R.}~\bibnamefont {Arita}}, \bibinfo {author} {\bibfnamefont {P.}~\bibnamefont {Wissgott}}, \bibinfo {author} {\bibfnamefont {A.}~\bibnamefont {Toschi}}, \bibinfo {author} {\bibfnamefont {H.}~\bibnamefont {Ikeda}}, \ and\ \bibinfo {author} {\bibfnamefont {K.}~\bibnamefont {Held}},\ }\href {\doibase https://doi.org/10.1016/j.cpc.2010.08.005} {\bibfield  {journal} {\bibinfo  {journal} {Computer Physics Communications}\ }\textbf {\bibinfo {volume} {181}},\ \bibinfo {pages} {1888} (\bibinfo {year} {2010})}\BibitemShut {NoStop}%
    \bibitem [{\citenamefont {Pizzi}\ \emph {et~al.}(2020)\citenamefont {Pizzi}, \citenamefont {Vitale}, \citenamefont {Arita}, \citenamefont {Blügel}, \citenamefont {Freimuth}, \citenamefont {G{\'{e}}ranton}, \citenamefont {Gibertini}, \citenamefont {Gresch}, \citenamefont {Johnson}, \citenamefont {Koretsune}, \citenamefont {Iba{\~{n}}ez-Azpiroz}, \citenamefont {Lee}, \citenamefont {Lihm}, \citenamefont {Marchand}, \citenamefont {Marrazzo}, \citenamefont {Mokrousov}, \citenamefont {Mustafa}, \citenamefont {Nohara}, \citenamefont {Nomura}, \citenamefont {Paulatto}, \citenamefont {Ponc{\'{e}}}, \citenamefont {Ponweiser}, \citenamefont {Qiao}, \citenamefont {Thöle}, \citenamefont {Tsirkin}, \citenamefont {Wierzbowska}, \citenamefont {Marzari}, \citenamefont {Vanderbilt}, \citenamefont {Souza}, \citenamefont {Mostofi},\ and\ \citenamefont {Yates}}]{Pizzi2020}%
      \BibitemOpen
      \bibfield  {author} {\bibinfo {author} {\bibfnamefont {G.}~\bibnamefont {Pizzi}}, \bibinfo {author} {\bibfnamefont {V.}~\bibnamefont {Vitale}}, \bibinfo {author} {\bibfnamefont {R.}~\bibnamefont {Arita}}, \bibinfo {author} {\bibfnamefont {S.}~\bibnamefont {Blügel}}, \bibinfo {author} {\bibfnamefont {F.}~\bibnamefont {Freimuth}}, \bibinfo {author} {\bibfnamefont {G.}~\bibnamefont {G{\'{e}}ranton}}, \bibinfo {author} {\bibfnamefont {M.}~\bibnamefont {Gibertini}}, \bibinfo {author} {\bibfnamefont {D.}~\bibnamefont {Gresch}}, \bibinfo {author} {\bibfnamefont {C.}~\bibnamefont {Johnson}}, \bibinfo {author} {\bibfnamefont {T.}~\bibnamefont {Koretsune}}, \bibinfo {author} {\bibfnamefont {J.}~\bibnamefont {Iba{\~{n}}ez-Azpiroz}}, \bibinfo {author} {\bibfnamefont {H.}~\bibnamefont {Lee}}, \bibinfo {author} {\bibfnamefont {J.-M.}\ \bibnamefont {Lihm}}, \bibinfo {author} {\bibfnamefont {D.}~\bibnamefont {Marchand}}, \bibinfo {author} {\bibfnamefont {A.}~\bibnamefont {Marrazzo}}, \bibinfo {author} {\bibfnamefont
      {Y.}~\bibnamefont {Mokrousov}}, \bibinfo {author} {\bibfnamefont {J.~I.}\ \bibnamefont {Mustafa}}, \bibinfo {author} {\bibfnamefont {Y.}~\bibnamefont {Nohara}}, \bibinfo {author} {\bibfnamefont {Y.}~\bibnamefont {Nomura}}, \bibinfo {author} {\bibfnamefont {L.}~\bibnamefont {Paulatto}}, \bibinfo {author} {\bibfnamefont {S.}~\bibnamefont {Ponc{\'{e}}}}, \bibinfo {author} {\bibfnamefont {T.}~\bibnamefont {Ponweiser}}, \bibinfo {author} {\bibfnamefont {J.}~\bibnamefont {Qiao}}, \bibinfo {author} {\bibfnamefont {F.}~\bibnamefont {Thöle}}, \bibinfo {author} {\bibfnamefont {S.~S.}\ \bibnamefont {Tsirkin}}, \bibinfo {author} {\bibfnamefont {M.}~\bibnamefont {Wierzbowska}}, \bibinfo {author} {\bibfnamefont {N.}~\bibnamefont {Marzari}}, \bibinfo {author} {\bibfnamefont {D.}~\bibnamefont {Vanderbilt}}, \bibinfo {author} {\bibfnamefont {I.}~\bibnamefont {Souza}}, \bibinfo {author} {\bibfnamefont {A.~A.}\ \bibnamefont {Mostofi}}, \ and\ \bibinfo {author} {\bibfnamefont {J.~R.}\ \bibnamefont {Yates}},\ }\href {\doibase
      10.1088/1361-648x/ab51ff} {\bibfield  {journal} {\bibinfo  {journal} {Journal of Physics: Condensed Matter}\ }\textbf {\bibinfo {volume} {32}},\ \bibinfo {pages} {165902} (\bibinfo {year} {2020})}\BibitemShut {NoStop}%
    \bibitem [{\citenamefont {Zhi}\ \emph {et~al.}(2022)\citenamefont {Zhi}, \citenamefont {Xu}, \citenamefont {Wu}, \citenamefont {Ning},\ and\ \citenamefont {Cao}}]{Zhi2022}%
      \BibitemOpen
      \bibfield  {author} {\bibinfo {author} {\bibfnamefont {G.-X.}\ \bibnamefont {Zhi}}, \bibinfo {author} {\bibfnamefont {C.}~\bibnamefont {Xu}}, \bibinfo {author} {\bibfnamefont {S.-Q.}\ \bibnamefont {Wu}}, \bibinfo {author} {\bibfnamefont {F.}~\bibnamefont {Ning}}, \ and\ \bibinfo {author} {\bibfnamefont {C.}~\bibnamefont {Cao}},\ }\href {\doibase https://doi.org/10.1016/j.cpc.2021.108196} {\bibfield  {journal} {\bibinfo  {journal} {Computer Physics Communications}\ }\textbf {\bibinfo {volume} {271}},\ \bibinfo {pages} {108196} (\bibinfo {year} {2022})}\BibitemShut {NoStop}%
    \end{thebibliography}

\begin{thebibliography}{11}%
  \makeatletter
  \providecommand \@ifxundefined [1]{%
   \@ifx{#1\undefined}
  }%
  \providecommand \@ifnum [1]{%
   \ifnum #1\expandafter \@firstoftwo
   \else \expandafter \@secondoftwo
   \fi
  }%
  \providecommand \@ifx [1]{%
   \ifx #1\expandafter \@firstoftwo
   \else \expandafter \@secondoftwo
   \fi
  }%
  \providecommand \natexlab [1]{#1}%
  \providecommand \enquote  [1]{``#1''}%
  \providecommand \bibnamefont  [1]{#1}%
  \providecommand \bibfnamefont [1]{#1}%
  \providecommand \citenamefont [1]{#1}%
  \providecommand \href@noop [0]{\@secondoftwo}%
  \providecommand \href [0]{\begingroup \@sanitize@url \@href}%
  \providecommand \@href[1]{\@@startlink{#1}\@@href}%
  \providecommand \@@href[1]{\endgroup#1\@@endlink}%
  \providecommand \@sanitize@url [0]{\catcode `\\12\catcode `\$12\catcode
    `\&12\catcode `\#12\catcode `\^12\catcode `\_12\catcode `\%12\relax}%
  \providecommand \@@startlink[1]{}%
  \providecommand \@@endlink[0]{}%
  \providecommand \url  [0]{\begingroup\@sanitize@url \@url }%
  \providecommand \@url [1]{\endgroup\@href {#1}{\urlprefix }}%
  \providecommand \urlprefix  [0]{URL }%
  \providecommand \Eprint [0]{\href }%
  \providecommand \doibase [0]{http://dx.doi.org/}%
  \providecommand \selectlanguage [0]{\@gobble}%
  \providecommand \bibinfo  [0]{\@secondoftwo}%
  \providecommand \bibfield  [0]{\@secondoftwo}%
  \providecommand \translation [1]{[#1]}%
  \providecommand \BibitemOpen [0]{}%
  \providecommand \bibitemStop [0]{}%
  \providecommand \bibitemNoStop [0]{.\EOS\space}%
  \providecommand \EOS [0]{\spacefactor3000\relax}%
  \providecommand \BibitemShut  [1]{\csname bibitem#1\endcsname}%
  \let\auto@bib@innerbib\@empty
  \bibitem [{bla()}]{blaha_2_S}%
    \BibitemOpen
    \href@noop {} {}\bibinfo {note} {P. Blaha, K. Schwarz, G. K. H. Madsen, D.
    Kvasnicka, J. Luitz, R. Laskowski, F. Tran, and L. D. Marks, \textit{WIEN2k,
    An Augmented Plane Wave + Local Orbitals Program for Calculating Crystal
    Properties} (Karlheinz Schwarz, Techn. Universit\"at Wien, Austria,
    2018).}\BibitemShut {Stop}%
  \bibitem [{\citenamefont {Czy\ifmmode~\dot{z}\else \.{z}\fi{}yk}\ and\
    \citenamefont {Sawatzky}(1994)}]{Czyzyk_LDA+U_AMF_S}%
    \BibitemOpen
    \bibfield  {author} {\bibinfo {author} {\bibfnamefont {M.~T.}\ \bibnamefont
    {Czy\ifmmode~\dot{z}\else \.{z}\fi{}yk}}\ and\ \bibinfo {author}
    {\bibfnamefont {G.~A.}\ \bibnamefont {Sawatzky}},\ }\href {\doibase
    10.1103/PhysRevB.49.14211} {\bibfield  {journal} {\bibinfo  {journal} {Phys.
    Rev. B}\ }\textbf {\bibinfo {volume} {49}},\ \bibinfo {pages} {14211}
    (\bibinfo {year} {1994})}\BibitemShut {NoStop}%
  \bibitem [{\citenamefont {Wigner}(1959)}]{Wigner_S}%
    \BibitemOpen
    \bibfield  {author} {\bibinfo {author} {\bibfnamefont {E.~P.}\ \bibnamefont
    {Wigner}},\ }\href@noop {} {\emph {\bibinfo {title} {Group Theory and Its
    Application to the Quantum Mechanics of Atomic Spectra}}}\ (\bibinfo
    {publisher} {Academic Press},\ \bibinfo {address} {New York},\ \bibinfo
    {year} {1959})\BibitemShut {NoStop}%
  \bibitem [{\citenamefont {Herring}(1937)}]{Herring1937_S}%
    \BibitemOpen
    \bibfield  {author} {\bibinfo {author} {\bibfnamefont {C.}~\bibnamefont
    {Herring}},\ }\href {\doibase 10.1103/PhysRev.52.361} {\bibfield  {journal}
    {\bibinfo  {journal} {Phys. Rev.}\ }\textbf {\bibinfo {volume} {52}},\
    \bibinfo {pages} {361} (\bibinfo {year} {1937})}\BibitemShut {NoStop}%
  \bibitem [{\citenamefont {Inui}\ \emph {et~al.}(1990)\citenamefont {Inui},
    \citenamefont {Tanabe},\ and\ \citenamefont
    {Onodera}}]{Inui-Tanabe-Onodera_S}%
    \BibitemOpen
    \bibfield  {author} {\bibinfo {author} {\bibfnamefont {T.}~\bibnamefont
    {Inui}}, \bibinfo {author} {\bibfnamefont {Y.}~\bibnamefont {Tanabe}}, \ and\
    \bibinfo {author} {\bibfnamefont {Y.}~\bibnamefont {Onodera}},\ }\href@noop
    {} {\emph {\bibinfo {title} {Group Theory and Its Applications in
    Physics}}},\ \bibinfo {series} {Springer Series in Solid-State Sciences},
    Vol.~\bibinfo {volume} {78}\ (\bibinfo  {publisher} {Springer-Verlag Berlin
    Heidelberg},\ \bibinfo {address} {Berlin, Heidelberg},\ \bibinfo {year}
    {1990})\BibitemShut {NoStop}%
  \bibitem [{\citenamefont {Bradley}\ and\ \citenamefont
    {Cracknell}(1972)}]{Bradley_S}%
    \BibitemOpen
    \bibfield  {author} {\bibinfo {author} {\bibfnamefont {C.~J.}\ \bibnamefont
    {Bradley}}\ and\ \bibinfo {author} {\bibfnamefont {A.~P.}\ \bibnamefont
    {Cracknell}},\ }\href@noop {} {\emph {\bibinfo {title} {The Mathematical
    Theory of Symmetry in Solids}}}\ (\bibinfo  {publisher} {Oxford University
    Press},\ \bibinfo {address} {Oxford},\ \bibinfo {year} {1972})\BibitemShut
    {NoStop}%
  \bibitem [{\citenamefont {Shiozaki}\ \emph {et~al.}(2022)\citenamefont
    {Shiozaki}, \citenamefont {Sato},\ and\ \citenamefont
    {Gomi}}]{Shiozaki2022_S}%
    \BibitemOpen
    \bibfield  {author} {\bibinfo {author} {\bibfnamefont {K.}~\bibnamefont
    {Shiozaki}}, \bibinfo {author} {\bibfnamefont {M.}~\bibnamefont {Sato}}, \
    and\ \bibinfo {author} {\bibfnamefont {K.}~\bibnamefont {Gomi}},\ }\href
    {\doibase 10.1103/PhysRevB.106.165103} {\bibfield  {journal} {\bibinfo
    {journal} {Phys. Rev. B}\ }\textbf {\bibinfo {volume} {106}},\ \bibinfo
    {pages} {165103} (\bibinfo {year} {2022})}\BibitemShut {NoStop}%
  \bibitem [{\citenamefont {Sumita}\ \emph {et~al.}(2019)\citenamefont {Sumita},
    \citenamefont {Nomoto}, \citenamefont {Shiozaki},\ and\ \citenamefont
    {Yanase}}]{Sumita2019_S}%
    \BibitemOpen
    \bibfield  {author} {\bibinfo {author} {\bibfnamefont {S.}~\bibnamefont
    {Sumita}}, \bibinfo {author} {\bibfnamefont {T.}~\bibnamefont {Nomoto}},
    \bibinfo {author} {\bibfnamefont {K.}~\bibnamefont {Shiozaki}}, \ and\
    \bibinfo {author} {\bibfnamefont {Y.}~\bibnamefont {Yanase}},\ }\href
    {\doibase 10.1103/PhysRevB.99.134513} {\bibfield  {journal} {\bibinfo
    {journal} {Phys. Rev. B}\ }\textbf {\bibinfo {volume} {99}},\ \bibinfo
    {pages} {134513} (\bibinfo {year} {2019})}\BibitemShut {NoStop}%
  \bibitem [{\citenamefont {Pizzi}\ \emph {et~al.}(2020)\citenamefont {Pizzi},
    \citenamefont {Vitale}, \citenamefont {Arita}, \citenamefont {Blügel},
    \citenamefont {Freimuth}, \citenamefont {G{\'{e}}ranton}, \citenamefont
    {Gibertini}, \citenamefont {Gresch}, \citenamefont {Johnson}, \citenamefont
    {Koretsune}, \citenamefont {Iba{\~{n}}ez-Azpiroz}, \citenamefont {Lee},
    \citenamefont {Lihm}, \citenamefont {Marchand}, \citenamefont {Marrazzo},
    \citenamefont {Mokrousov}, \citenamefont {Mustafa}, \citenamefont {Nohara},
    \citenamefont {Nomura}, \citenamefont {Paulatto}, \citenamefont
    {Ponc{\'{e}}}, \citenamefont {Ponweiser}, \citenamefont {Qiao}, \citenamefont
    {Thöle}, \citenamefont {Tsirkin}, \citenamefont {Wierzbowska}, \citenamefont
    {Marzari}, \citenamefont {Vanderbilt}, \citenamefont {Souza}, \citenamefont
    {Mostofi},\ and\ \citenamefont {Yates}}]{Pizzi2020}%
    \BibitemOpen
    \bibfield  {author} {\bibinfo {author} {\bibfnamefont {G.}~\bibnamefont
    {Pizzi}}, \bibinfo {author} {\bibfnamefont {V.}~\bibnamefont {Vitale}},
    \bibinfo {author} {\bibfnamefont {R.}~\bibnamefont {Arita}}, \bibinfo
    {author} {\bibfnamefont {S.}~\bibnamefont {Blügel}}, \bibinfo {author}
    {\bibfnamefont {F.}~\bibnamefont {Freimuth}}, \bibinfo {author}
    {\bibfnamefont {G.}~\bibnamefont {G{\'{e}}ranton}}, \bibinfo {author}
    {\bibfnamefont {M.}~\bibnamefont {Gibertini}}, \bibinfo {author}
    {\bibfnamefont {D.}~\bibnamefont {Gresch}}, \bibinfo {author} {\bibfnamefont
    {C.}~\bibnamefont {Johnson}}, \bibinfo {author} {\bibfnamefont
    {T.}~\bibnamefont {Koretsune}}, \bibinfo {author} {\bibfnamefont
    {J.}~\bibnamefont {Iba{\~{n}}ez-Azpiroz}}, \bibinfo {author} {\bibfnamefont
    {H.}~\bibnamefont {Lee}}, \bibinfo {author} {\bibfnamefont {J.-M.}\
    \bibnamefont {Lihm}}, \bibinfo {author} {\bibfnamefont {D.}~\bibnamefont
    {Marchand}}, \bibinfo {author} {\bibfnamefont {A.}~\bibnamefont {Marrazzo}},
    \bibinfo {author} {\bibfnamefont {Y.}~\bibnamefont {Mokrousov}}, \bibinfo
    {author} {\bibfnamefont {J.~I.}\ \bibnamefont {Mustafa}}, \bibinfo {author}
    {\bibfnamefont {Y.}~\bibnamefont {Nohara}}, \bibinfo {author} {\bibfnamefont
    {Y.}~\bibnamefont {Nomura}}, \bibinfo {author} {\bibfnamefont
    {L.}~\bibnamefont {Paulatto}}, \bibinfo {author} {\bibfnamefont
    {S.}~\bibnamefont {Ponc{\'{e}}}}, \bibinfo {author} {\bibfnamefont
    {T.}~\bibnamefont {Ponweiser}}, \bibinfo {author} {\bibfnamefont
    {J.}~\bibnamefont {Qiao}}, \bibinfo {author} {\bibfnamefont {F.}~\bibnamefont
    {Thöle}}, \bibinfo {author} {\bibfnamefont {S.~S.}\ \bibnamefont {Tsirkin}},
    \bibinfo {author} {\bibfnamefont {M.}~\bibnamefont {Wierzbowska}}, \bibinfo
    {author} {\bibfnamefont {N.}~\bibnamefont {Marzari}}, \bibinfo {author}
    {\bibfnamefont {D.}~\bibnamefont {Vanderbilt}}, \bibinfo {author}
    {\bibfnamefont {I.}~\bibnamefont {Souza}}, \bibinfo {author} {\bibfnamefont
    {A.~A.}\ \bibnamefont {Mostofi}}, \ and\ \bibinfo {author} {\bibfnamefont
    {J.~R.}\ \bibnamefont {Yates}},\ }\href {\doibase 10.1088/1361-648x/ab51ff}
    {\bibfield  {journal} {\bibinfo  {journal} {Journal of Physics: Condensed
    Matter}\ }\textbf {\bibinfo {volume} {32}},\ \bibinfo {pages} {165902}
    (\bibinfo {year} {2020})}\BibitemShut {NoStop}%
  \bibitem [{\citenamefont {Kuneš}\ \emph {et~al.}(2010)\citenamefont {Kuneš},
    \citenamefont {Arita}, \citenamefont {Wissgott}, \citenamefont {Toschi},
    \citenamefont {Ikeda},\ and\ \citenamefont {Held}}]{Kunes2010}%
    \BibitemOpen
    \bibfield  {author} {\bibinfo {author} {\bibfnamefont {J.}~\bibnamefont
    {Kuneš}}, \bibinfo {author} {\bibfnamefont {R.}~\bibnamefont {Arita}},
    \bibinfo {author} {\bibfnamefont {P.}~\bibnamefont {Wissgott}}, \bibinfo
    {author} {\bibfnamefont {A.}~\bibnamefont {Toschi}}, \bibinfo {author}
    {\bibfnamefont {H.}~\bibnamefont {Ikeda}}, \ and\ \bibinfo {author}
    {\bibfnamefont {K.}~\bibnamefont {Held}},\ }\href {\doibase
    https://doi.org/10.1016/j.cpc.2010.08.005} {\bibfield  {journal} {\bibinfo
    {journal} {Computer Physics Communications}\ }\textbf {\bibinfo {volume}
    {181}},\ \bibinfo {pages} {1888} (\bibinfo {year} {2010})}\BibitemShut
    {NoStop}%
  \bibitem [{\citenamefont {Zhi}\ \emph {et~al.}(2022)\citenamefont {Zhi},
    \citenamefont {Xu}, \citenamefont {Wu}, \citenamefont {Ning},\ and\
    \citenamefont {Cao}}]{Zhi2022}%
    \BibitemOpen
    \bibfield  {author} {\bibinfo {author} {\bibfnamefont {G.-X.}\ \bibnamefont
    {Zhi}}, \bibinfo {author} {\bibfnamefont {C.}~\bibnamefont {Xu}}, \bibinfo
    {author} {\bibfnamefont {S.-Q.}\ \bibnamefont {Wu}}, \bibinfo {author}
    {\bibfnamefont {F.}~\bibnamefont {Ning}}, \ and\ \bibinfo {author}
    {\bibfnamefont {C.}~\bibnamefont {Cao}},\ }\href {\doibase
    https://doi.org/10.1016/j.cpc.2021.108196} {\bibfield  {journal} {\bibinfo
    {journal} {Computer Physics Communications}\ }\textbf {\bibinfo {volume}
    {271}},\ \bibinfo {pages} {108196} (\bibinfo {year} {2022})}\BibitemShut
    {NoStop}%
  \end{thebibliography}
%

    \clearpage

    \renewcommand{\thesection}{S\arabic{section}}

    \setcounter{section}{0}
    
    \renewcommand{\theequation}{S\arabic{equation}}
    
    \setcounter{equation}{0}
    
    \renewcommand{\thefigure}{S\arabic{figure}}
    
    \setcounter{figure}{0}
    
    \renewcommand{\thetable}{S\arabic{table}}
    
    \setcounter{table}{0}
    
    \renewcommand*{\citenumfont}[1]{S#1}
    
    \renewcommand*{\bibnumfmt}[1]{[S#1]}
    
    \makeatletter
    
    \c@secnumdepth = 2
    
    \makeatother
    
  
    \begin{center}

    {\large \textmd{Supplemental Materials:} \\[0.3em]
    
    {\bfseries Correlation-induced Fermi surface evolution and topological crystalline superconductivity in CeRh$_2$As$_2$}}
    
    \end{center}

    \section{Details of band calculation}

    We conduct the DFT calculations for paramagnetic CeRh$_2$As$_2$ using the \wien package \cite{blaha_2_S}. We use the full-potential linearized augmented plane wave $+$ local orbitals method within the generalized gradient approximation. The spin-orbit coupling is included in all calculations. 
    We set the muffin-tin radii ($R_{\rm MT}$) of $2.5$, $2.34$, and $2.23$ a.u. for Ce, Rh, and As, respectively.
    The maximum reciprocal lattice vector $K_{\rm max}$ is given by $R_{\rm MT}K_{\rm max}=8$.
    The Brillouin zone sampling is performed with a $16\times16\times7$ $k$-mesh.
    For DFT$+U$, We set Hund's coupling $J=0$ and subtract the double-counting correlation by the around mean-field formula \cite{Czyzyk_LDA+U_AMF_S}. 
    
    \section{Classification of superconducting gap nodes}
    
    \subsection{Classification on the
    \texorpdfstring{$k_x+k_y=0$}{kx+ky=0} lines with \texorpdfstring{$k_z=0$}{kz=0} and \texorpdfstring{$\pi$}{pi}}
    First, we focus on the $k_x+k_y=0$ line with $k_z=0$ and $\pi$. On these lines, the operators' commutation relation between the glide symmetry $G$ and the inversion symmetry ${\cal I}=\{I |\bm 0\}$ is well defined.
    The magnetic little cogroup is given by
    \begin{equation}
     \bar{\mathfrak{G}}^{\bm{k}} = \bar{\cal G}^{\bm{k}} + \mathfrak{T} \bar{\cal G}^{\bm{k}} + \mathfrak{C} \bar{\cal G}^{\bm{k}} + \Gamma \bar{\cal G}^{\bm{k}},
    \end{equation}
    where $\bar{\cal G}^{\bm{k}}$ is the unitary little cogroup.
    On these high symmetry lines, $\bar{\cal G}^{\bm{k}}=\{E, G\}$ isomorphic to the $C_s$ point group.
    Here, we defined a time-reversal symmetry like operator, $\mathfrak{T} \equiv {\cal T}S_{2a}$, where ${\cal T}$ is the time-reversal symmetry and $S_{2a} \equiv \{2_{110}  |  \bm a/2 + \bm b/2 \}$ is the screw symmetry. We also defined a particle-hole symmetry like operator  $\mathfrak{C} \equiv {\cal C I}$ with ${\cal C}$ being the particle-hole symmetry, and accordingly a chiral symmetry like operator is $\Gamma \equiv \mathfrak{T} \mathfrak{C}$.
    Note that ${\cal TI}$ is not preserved in the presence of the magnetic field.
    Then, using the factor system $\{z_{g, h}^{\bm{k}}\} \in Z^2(\bar{\mathfrak{G}}^{\bm{k}}, \mathrm{U}(1)_\phi)$, we execute the Wigner criteria~\cite{Wigner_S, Herring1937_S, Inui-Tanabe-Onodera_S, Bradley_S, Shiozaki2022_S} for $\mathfrak{T}$ and $\mathfrak{C}$, and the orthogonality test~\cite{Inui-Tanabe-Onodera_S, Shiozaki2022_S} for $\Gamma$:
    \begin{align}
     W_\alpha^{\mathfrak{T}} &\equiv \frac{1}{|\bar{\cal G}^{\bm{k}}|} \sum_{g \in \bar{\cal G}^{\bm{k}}} z_{\mathfrak{T} g, \mathfrak{T} g}^{\bm{k}} \chi[\bar{\gamma}^{\bm{k}}_\alpha((\mathfrak{T} g)^2)] =
     \begin{cases}
      1, \\
      -1, \\
      0,
     \end{cases} \label{eq:Wigner_criterion_T} \\
     W_\alpha^{\mathfrak{C}} &\equiv \frac{1}{|\bar{\cal G}^{\bm{k}}|} \sum_{g \in \bar{\cal G}^{\bm{k}}} z_{\mathfrak{C} g, \mathfrak{C} g}^{\bm{k}} \chi[\bar{\gamma}^{\bm{k}}_\alpha((\mathfrak{C} g)^2)] =
     \begin{cases}
      1, \\
      -1, \\
      0,
     \end{cases} \label{eq:Wigner_criterion_C} \\
     W_\alpha^\Gamma &\equiv \frac{1}{|\bar{\cal G}^{\bm{k}}|} \sum_{g \in \bar{\cal G}^{\bm{k}}} \frac{z_{g, \Gamma}^{\bm{k} *}}{z_{\Gamma, \Gamma^{-1} g \Gamma}^{\bm{k} *}} \chi[\bar{\gamma}^{\bm{k}}_\alpha(\Gamma^{-1} g \Gamma)^*] \chi[\bar{\gamma}^{\bm{k}}_\alpha(g)] =
     \begin{cases}
      1, \\
      0,
     \end{cases} \label{eq:orthogonality_test_G}
    \end{align}
    where $\chi$ is the character of the representation and $\bar{\gamma}^{\bm{k}}_\alpha$ is the projective irreducible representation of $\bar{\cal G}^{\bm{k}}$ with the total angular momentum $\alpha$ of the normal Bloch state. By these tests, we can identify the effective Altland-Zirnbauer (EAZ) symmetry class of the Bogoliubov-de Gennes Hamiltonian on the high symmetry points. 
    
    The projective irreducible representation $\bar{\gamma}^{\bm{k}}_\alpha$ of $\bar{\cal G}^{\bm{k}}$ is given by
    \begin{equation}
        \bar{\gamma}^{\bm{k}}_{\pm1/2}(E) = 1, \quad
        \bar{\gamma}^{\bm{k}}_{\pm1/2}(G) = \pm i,
    \end{equation}
    which corresponds to the Bloch state $c^\dag_{\bm k, \pm1/2}$ for spin-up ($+1/2$) and spin-down ($-1/2$) with the spin-orbit coupling. By the above tests, we investigate how the irreducible representation $\bar{\gamma}^{\bm{k}}_\alpha$ pairs up with the partner under the additional symmetry $a_0 = {\mathfrak{T}}$, ${\mathfrak{C}}$, and $\Gamma$. In other words, we can clarify the relation between $c^\dag_{\bm k, \alpha}$ and $a_0 c^\dag_{\bm k, \alpha}a_0^{-1}$.

    Now, we prepare some algebraic relations between these operators.
    We first focus on the $A_u$ pair wave function in the $C_{2h}$ point group on these high symmetry lines: the compatibility relation is given by $A_u\uparrow C_{4h}=A_u+B_u$, and thus, the PDW state corresponds to this case.
    For the screw operator $S_{2a}$ and the glide operator $G$, we obtain
    
    \begin{equation}
        S_{2a}^2 = \{-E | \bm a + \bm b\}, \quad G^2 = \{-E | \bm a + \bm b\},
    \end{equation}
    and $\{S_{2a}, G\} = 0$.
    Note that $\{E | \bm a + \bm b\}$ is reduced to a phase factor $\exp[-i(k_x+k_y)]$, which gives the factor 1 for $k_x+k_y=0$, and then equivalent to the identity operator.
    By ${\cal T}^2=\{-E | \bm 0\}$ and $[{\cal T}, S_{2a}]=[{\cal T}, G]=0$, it follows that
    \begin{equation}
        (\mathfrak{T}E)^2 = \{E | \bm a + \bm b\}, \quad
        (\mathfrak{T}G)^2 = \{E | 2\bm a + 2\bm b\}.
    \end{equation}
    For the particle-hole operator ${\cal C}$ and the inversion operator ${\cal I}$, we get
    \begin{equation}
        \{{\cal C}, {\cal I}\}=0, \quad \mathfrak{C}^2=\{-E | \bm 0\},
    \end{equation}
    in the odd-parity superconducting states. For $\cal C$ and $G$, we have $\{{\cal C}, G\}=0$, since $A_u$ is the mirror-odd representation.
    By ${\cal I} G=G {\cal I}\{E  |  \bm a + \bm b\}$, it indicates that
    \begin{equation}
        (\mathfrak{C}G)^2 = \{-E|\bm 0\}.
    \end{equation}
    
    By the above algebraic relations, the Wigner criteria and the orthogonality test result in 
    $(W_{+1/2}^{\mathfrak{T}}, W_{+1/2}^{\mathfrak{C}}, W_{+1/2}^{\Gamma})=(1,-1,1)$. 
    Therefore, the system is identified as the EAZ symmetry class CI. 
    Since the class CI is classified into $0$ (trivial) referring to the knowledge of the \textit{K} theory, the gap opens on the $k_x+k_y=0$ line by the $A_u$ ($A_u$ or $B_u$ in $C_{4h}$) pair wave function when the FSs of normal state cross the line (see Ref.~\cite{Sumita2019_S} for detail).
    
    Next we focus on the $B_u$ pair wave function in the $C_{2h}$ point group: the compatibility relation is $B_u\uparrow C_{4h}=2E_u$. 
    The Wigner criteria and the orthogonality test yield $(W_{+1/2}^{\mathfrak{T}}, W_{+1/2}^{\mathfrak{C}}, W_{+1/2}^{\Gamma})=(1,0,0)$. Here, we used $[{\cal C}, G]=0$ and $(\mathfrak{C}G)^2 = \{E|\bm 0\}$ since $B_u$ is the mirror-even representation. Therefore, the system is identified as the EAZ symmetry class AI. Since the class AI is classified into $\mathbb{Z}$, the gap closes on the $k_x+k_y=0$ line by the $B_u$ ($E_u$ in $C_{4h}$) pair wave function.
    
    \subsection{Classification on the
    \texorpdfstring{$k_x=k_y=0$}{kx=ky=0} line (\texorpdfstring{$k_z$}{kz} axis)}
    \label{ap:Gap_kz}
    By defining $\mathfrak{T} \equiv {\cal T}C_{2x}$, where $C_{2x} \equiv \{2_{100}  |  \bm 0 \}$ is rotational symmetry,
    the magnetic little cogroup is given by
    \begin{equation}
     \bar{\mathfrak{G}}^{\bm{k}} = \bar{\cal G}^{\bm{k}} + \mathfrak{T} \bar{\cal G}^{\bm{k}} + \mathfrak{C} \bar{\cal G}^{\bm{k}} + \Gamma \bar{\cal G}^{\bm{k}}.
    \end{equation}
    The unitary little cogroup is $\bar{\cal G}^{\bm{k}}=\{E, {\tilde C}_{4z}, ({\tilde C}_{4z})^2, ({\tilde C}_{4z})^3\}$ isomorphic to the $C_4$ point group, where ${\tilde C}_{4z} \equiv \{4_{001}  |  \bm a/2 \}$ and $({\tilde C}_{4z})^2=\{2_{001}  |  \bm a/2 + \bm b/2\} \equiv {\tilde C}_{2z}$.
    Since we are focusing on $k_x=k_y=0$, a half translation in the $xy$ plane is not an essential factor for the following classification.
    The irreducible representation is given by
    \begin{align}    \bar{\gamma}^{\bm{k}}_{+1/2}(E) &= 1, \quad \bar{\gamma}^{\bm{k}}_{+1/2}({\tilde C}_{4z}) = e^{+\pi i/4}, \\
        \bar{\gamma}^{\bm{k}}_{+1/2}({\tilde C}_{2z}) &= +i, \quad
        \bar{\gamma}^{\bm{k}}_{+1/2}(({\tilde C}_{4z})^3) = e^{+3\pi i/4}.
    \end{align}
    The algebraic relations for these operators are obtained as
    \begin{align}
        C_{2x} {\tilde C}_{4z} &= \{ 2_{1\bar 1 0} | {\bm a}/2\} \equiv C_{2b}, \\
        \{ C_{2x}, {\tilde C}_{2z}\} &= \{ C_{2b}, {\tilde C}_{2z}\} = 0.
    \end{align}
    Here, the commutation relation between $C_{2x}$ and ${\tilde C}_{4z}$ is
    \begin{equation}
        C_{2x} {\tilde C}_{4z} C_{2z} = {\tilde C}_{4z} C_{2x}, \quad C_{2z} \equiv \{2_{001}|\bm 0\}.
    \end{equation}
    Then, we get the Wigner criterion for $\mathfrak{T}$:
    \begin{align}
     W_{+1/2}^{\mathfrak{T}} & = \frac{1}{4} \{
     z_{\mathfrak{T} E, \mathfrak{T} E}^{\bm{k}} \chi[\bar{\gamma}^{\bm{k}}_\alpha((\mathfrak{T} E)^2)] \nonumber \\
     & + z_{\mathfrak{T} {\tilde C}_{4z}, \mathfrak{T} {\tilde C}_{4z}}^{\bm{k}} \chi[\bar{\gamma}^{\bm{k}}_\alpha((\mathfrak{T} {\tilde C}_{4z})^2)] \nonumber \\
     & + z_{\mathfrak{T} {\tilde C}_{2z}, \mathfrak{T} {\tilde C}_{2z}}^{\bm{k}} \chi[\bar{\gamma}^{\bm{k}}_\alpha((\mathfrak{T} {\tilde C}_{2z})^2)] \nonumber \\
     & + z_{\mathfrak{T} {\tilde C}_{4z}^3, \mathfrak{T} {\tilde C}_{4z}^3}^{\bm{k}} \chi[\bar{\gamma}^{\bm{k}}_\alpha((\mathfrak{T} {\tilde C}_{4z}^3)^2)]\} \nonumber \\
     & = \frac{1}{4}\{1+1+1+1\} = 1.
    \end{align}
    Next, we execute the tests for $\mathfrak{C}$ and $\Gamma$, restricting the pairing symmetry to the $A_u$ state.
    By some relations:
    \begin{equation}
        [{\cal C}, {\tilde C}_{4z}] = [{\cal C}, {\tilde C}_{2z}] = 0, \quad
        {\cal I} {\tilde C}_{4z} = \{E | -{\bm a} \} {\tilde C}_{4z} {\cal I},
    \end{equation}
    we obtain
    \begin{align}
        & (\mathfrak{C} {\tilde C}_{4z})^2 = \{ -E | -{\bm a} \} {\tilde C}_{2z}, \\
        & (\mathfrak{C} {\tilde C}_{2z})^2 = \{ E | {\bm a} + {\bm b}\}, \\
        & \Gamma^{-1} {\tilde C}_{4z} \Gamma = \{ E | -{\bm a}/2 - {\bm b}/2 \} ({\tilde C}_{4z})^3.
    \end{align}
    The Wigner criterion and the orthogonality test result in
    \begin{align}
     W_{+1/2}^{\mathfrak{C}} & = \frac{1}{4} \{-1 -i + 1 + i\}  = 0, \\
     W_{+1/2}^{\Gamma} & = \frac{1}{4} \{1 + [(e^{+i 3\pi /4})^*] (e^{+i \pi /4}) \nonumber \\
     & - (i^*) i + [(e^{+i \pi /4})^*] (e^{+i 3\pi /4})
        \}  = 0.
    \end{align}
    Thus, a set of $(W_{+1/2}^{\mathfrak{T}}, W_{+1/2}^{\mathfrak{C}} ,W_{+1/2}^{\Gamma})=(1, 0, 0)$ identifies the system with the EAZ symmetry class AI; the gap closes on the $k_z$ axis by the $A_u$ pair wave function. 
    
    Next, we assume the $B_u$ pairing state, which has $C_4$-odd and $C_2$-even symmetry: $\{ {\cal C}, {\tilde C}_{4z} \} = 0$ and $[{\cal C}, {\tilde C}_{2z}] = 0$.
    The Wigner criterion for ${\mathfrak C}$ and the orthogonality test for $\Gamma$ are given by
    \begin{align}
     W_{+1/2}^{\mathfrak{C}} & = \frac{1}{4} \{-1 +i + 1 - i\}  = 0, \\
     W_{+1/2}^{\Gamma} & = \frac{1}{4} \{1 - [(e^{+i 3\pi /4})^*] (e^{+i \pi /4}) \nonumber \\
     & - (i^*) i - [(e^{+i \pi /4})^*] (e^{+i 3\pi /4})
        \}  = 0.
    \end{align}
    Thus, the EAZ symmetry is class AI. The gap closes on the $k_z$ axis by the $B_u$ pair wave function.
    
    The $E^{1,2}_{u}$ states satisfy the following commutation relations (ignoring the phase factor):
    \begin{align}
        & {\tilde C}_{4z} {\cal C}  = \mp i {\cal C} {\tilde C}_{4z}, \quad 
        \{ {\cal C}, {\tilde C}_{2z} \} = 0, \\
        & (\mathfrak{C} {\tilde C}_{4z})^2 = \mp i {\tilde C}_{2z}, \quad
        (\mathfrak{C} {\tilde C}_{4z}^3)^2 = \mp i {\tilde C}_{2z}, \\
        & \Gamma^{-1} {\tilde C}_{4z} \Gamma = \mp i {\tilde C}_{4z}^3, \quad
        \Gamma^{-1} {\tilde C}_{4z}^3 \Gamma = \pm i {\tilde C}_{4z},
    \end{align}
    where the upper (lower) sign corresponds to the $E^{1}_{u}$ ($E^{2}_{u}$) state.
    Therefore, the Wigner criterion for ${\mathfrak C}$ and the orthogonality test for $\Gamma$ are obtained as
    \begin{align}
     W_{+1/2}^{\mathfrak{C}} & = \frac{1}{4} \{-1 - i(i) - 1 - i (i)\}  = 0, \\
     W_{+1/2}^{\Gamma} & = \frac{1}{4} \{1 - i [(e^{+i 3\pi /4})^*] (e^{+i \pi /4}) \nonumber \\
     & + (i^*) i - i [(e^{+i \pi /4})^*] (e^{+i 3\pi /4})
        \}  = 0,
    \end{align}
    for the $E^{1}_{u}$ pairing state, while 
    \begin{align}
     W_{+1/2}^{\mathfrak{C}} & = \frac{1}{4} \{-1 + i(i) - 1 + i (i)\}  = -1, \\
     W_{+1/2}^{\Gamma} & = \frac{1}{4} \{1 + i [(e^{+i 3\pi /4})^*] (e^{+i \pi /4}) \nonumber \\
     & + (i^*) i - i [(e^{+i \pi /4})^*] (e^{+i 3\pi /4})
        \}  = 1, 
    \end{align}
    for the $E^{2}_{u}$ pairing state. The above results are summarized in Table III in the main text.
    
    \section{Majorana surface states}
\begin{figure}[tbp]
    \includegraphics[width=1.0\linewidth]{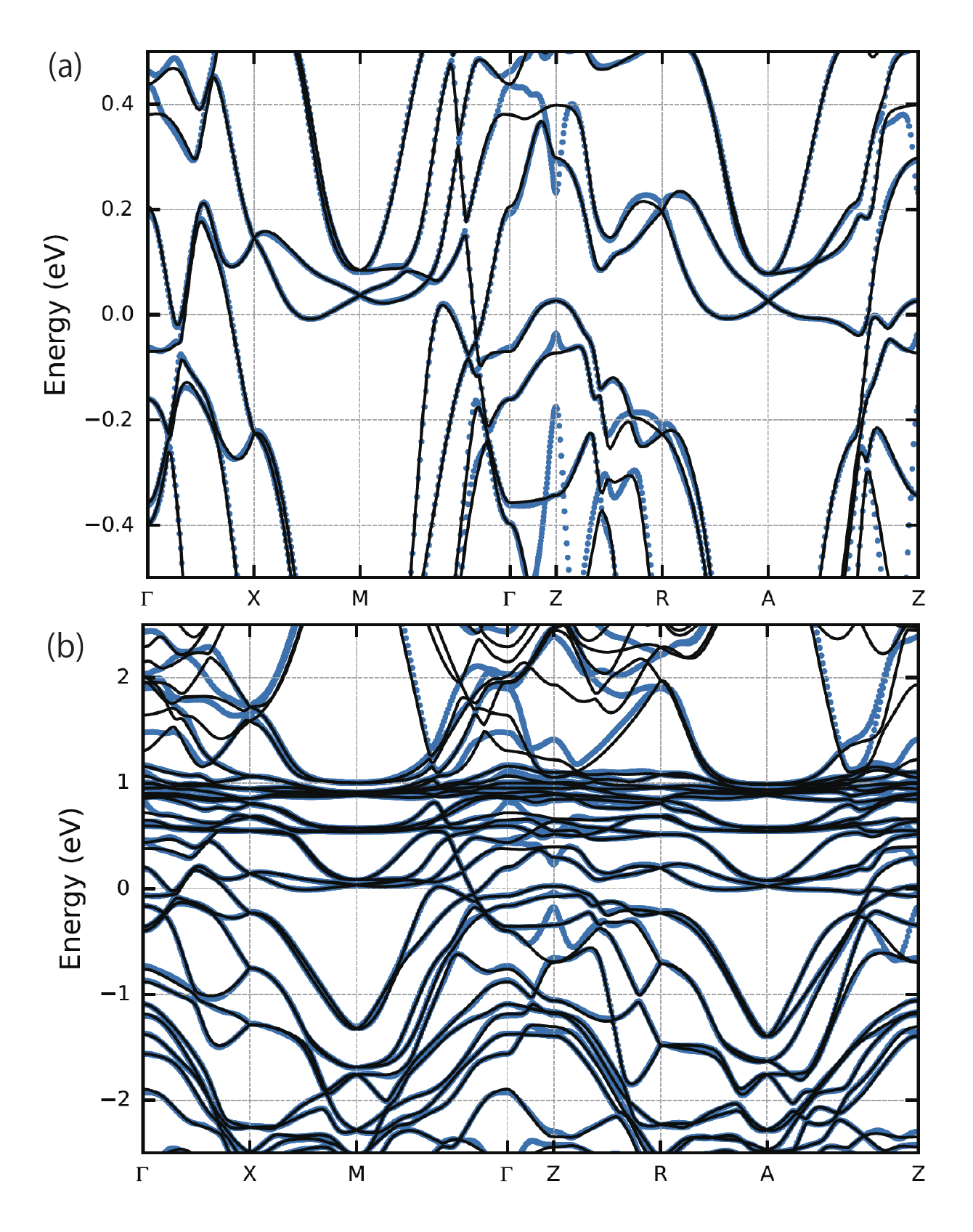}
    \centering
    \caption{(a)-(b) Comparison of DFT$+U$ and Wannier model. Black lines: Band structure obtained by the DFT$+U$ calculation for $U=2.0$ eV along the high symmetry lines. Blue marks: Band structure of the Wannier model.
    \label{fig:model}}
    \end{figure}

    To demonstrate the emergence of Majorana surface states, we construct the {\it ab} {\it initio} Wannier model by using \textsc{wannier90} \cite{Pizzi2020} through \textsc{wien2wannier} \cite{Kunes2010} for $U=2.0$ eV.
    Projectors from the Bloch states to the Wannier functions consist of 4$f$ orbitals on the two Ce sites, 4$d$ orbitals on the four Rh sites, and 4$p$ orbitals on the four As sites with spin degree of freedom, resulting in a 92-band model.
    We perform a one-shot Wannierization with $8 \times 8 \times 8$ $\bm k$-points.
    To reduce numerical errors, we symmetrized the hopping integrals by symmetry operations acting on the Wannier basis through WannSymm code \cite{Zhi2022}.
    The kinetic part of the Hamiltonian is given by
    \begin{align}
        \mathcal{\hat{H}}_{\rm kin} &= \sum_{{\bm k},\alpha,\beta} \hat{\mathcal{H}}_{\rm kin}(\bm{k}) c^\dag_{{\bm k}\alpha} c_{{\bm k}\beta}, \\
        \hat{\mathcal{H}}_{\rm kin}(\bm{k}) &= \sum_{\bm R} e^{i\bm k \cdot \bm R} t^{\alpha\beta}_{{\bm i}{\bm j}}
    \end{align}
    where $t^{\alpha\beta}_{{\bm i}{\bm j}}$ is the hopping integrals, $\bm R = \bm i - \bm j$, and $\alpha=(m,l,s)$ denotes the internal degrees of freedom of sublattice $m$, orbital $l$, and spin $s$.
    The energy eigenvalues are obtained by diagonalizing the matrix representation of the kinetic Hamiltonian $\mathcal{H}_{\rm kin}(\bm{k})$.
    The band structure of the constructed Wannier model is in good agreement with the DFT$+U$ calculation (Fig. \ref{fig:model}).
    
    Next, we construct a Bogoliubov-de Gennes (BdG) Hamiltonian using the Wannier model.
    The normal part of the Hamiltonian is given by
    \begin{align}
        \mathcal{\hat{H}} &= \mathcal{\hat{H}}_{\rm kin} + \mathcal{\hat{H}}_{\rm Z} 
        = \sum_{\boldsymbol{k}} \hat{C}_{\boldsymbol{k}}^{\dagger} \hat{\mathcal{H}}(\boldsymbol{k}) \hat{C}_{\boldsymbol{k}},\\
        \hat{C}_{\boldsymbol{k}} &= (
        c_{\boldsymbol{k} {\rm Ce_1} f_{z^3} \uparrow}, 
        c_{\boldsymbol{k} {\rm Ce_1} f_{z^3} \downarrow}, 
        c_{\boldsymbol{k} {\rm Ce_1} f_{xz^2} \uparrow},
        c_{\boldsymbol{k} {\rm Ce_1} f_{xz^2} \downarrow}, \nonumber \\ 
        &\quad \ \cdots,
        c_{\boldsymbol{k} {\rm As_2} p_{y} \uparrow},
        c_{\boldsymbol{k} {\rm As_2} p_{y} \downarrow}
        )^{\top},
    \end{align}
    where $\hat{C}_{\boldsymbol{k}}$ is a 92-dimensional vector operator, and $\mathcal{\hat{H}}_{\rm Z}$ is the Zeeman coupling term
    \begin{align}
        \mathcal{\hat{H}}_{\rm Z} = -g\mu_{\mathrm{B}} \sum_{k, m, l, \sigma, \sigma^{\prime}} \boldsymbol{H} \cdot \boldsymbol{\sigma}_{s s^{\prime}} c_{\boldsymbol{k} m l s}^{\dagger} c_{\boldsymbol{k} m l s^{\prime}},
    \end{align}
    with the $g$-factor $g$ and Pauli matrices $\hat{{\bm \sigma}}=(\hat{\sigma}_x, \hat{\sigma}_y, \hat{\sigma}_z)$ for spin space. Hereafter we assume $g=10$ and $\boldsymbol{H}=(0,0,H)$. In the Nambu space, the BdG Hamiltonian takes the form:
    \begin{align}
    \mathcal{\hat{H}}_{\mathrm{BdG}} &= \frac{1}{2} \sum_{\boldsymbol{k}} \hat{\Phi}_{\boldsymbol{k}}^{\dagger} \hat{\mathcal{H}}_{\mathrm{BdG}}(\boldsymbol{k}) \hat{\Phi}_{\boldsymbol{k}},\\
    \hat{\mathcal{H}}_{\mathrm{BdG}}(\boldsymbol{k}) &= \left(\begin{array}{cc}
    \hat{\mathcal{H}}(\boldsymbol{k}) & \hat{\Delta}(\boldsymbol{k}) \\
    \hat{\Delta}^{\dagger}(\boldsymbol{k}) & -\hat{\mathcal{H}}^{\top}(-\boldsymbol{k})
    \end{array}\right),
    \end{align}
    where $\hat{\Phi}^{\dag}_{\boldsymbol{k}}=(\hat{C}^{\dag}_{\boldsymbol{k}}, \hat{C}^{\top}_{-\boldsymbol{k}})$ is a 184-dimensional vector operator.
    
    We now introduce the order parameter $\hat{\Delta}(\boldsymbol{k})$ of the odd-parity superconducting state in the PDW phase.
    For simplicity, we assume that only the gap functions on the Ce atoms are finite:
    \begin{align}
        \hat{\Delta}(\boldsymbol{k}) &= 
        \left(\begin{array}{cc}
        \hat{\Delta}_{\rm Ce}(\boldsymbol{k}) & \\
         & \hat{\bm 0}_{64}
        \end{array}\right),\\
    \end{align}
    where $\hat{\bm 0}_{64}$ is the $64\times64$ zero matrix in Rh and As subspace.
    In addition, we ignore interorbital and intersublattice components and the orbital-dependence in the gap function. 
    By this simplification, the gap function is given by
    \begin{align}
        \hat{\Delta}_{\rm Ce}(\boldsymbol{k}) = \left( \psi_0(\boldsymbol{k}) \hat{\sigma}_{0} i\hat{\sigma}_{y} \otimes \hat{\tau}_{z} + \boldsymbol{d}(\boldsymbol{k}) \cdot \hat{\boldsymbol{\sigma}}i\hat{\sigma}_{y} \otimes \hat{\tau}_{0} \right) \otimes \hat{1}_{7},
    \end{align}
    where $\hat{\sigma}_{0}$ and $\hat{\tau}_{0}$ are the $2\times2$ identity matrix for spin and sublattice space, $\hat{1}_7$ is the $7\times7$ identity matrix for orbital space, and $\hat{{\bm \tau}}=(\hat{\tau}_x, \hat{\tau}_y, \hat{\tau}_z)$ are Pauli matrices for sublattice space. $\psi_0(\boldsymbol{k})$ is a basis function for the spin-singlet pairing component, and $\boldsymbol{d}(\boldsymbol{k})$ is a $d$-vector for the spin-triplet pairing component.
    For the $A_u$ pairing state, we have
    \begin{equation}
        \boldsymbol{d}(\boldsymbol{k}) = a_1(k_x \hat{{\bm x}} + k_y \hat{{\bm y}}) + a_2 k_z \hat{{\bm z}} + a_3(k_y \hat{\bm{x}} - k_x \hat{\bm{y}}),
    \end{equation}
    and
    \begin{equation}    
        \psi_0(\boldsymbol{k}) = a_4 [1+k_xk_y(k_x^2-k_y^2)].
    \end{equation}
    
\begin{figure}[tbp]
    \includegraphics[width=1.0\linewidth]{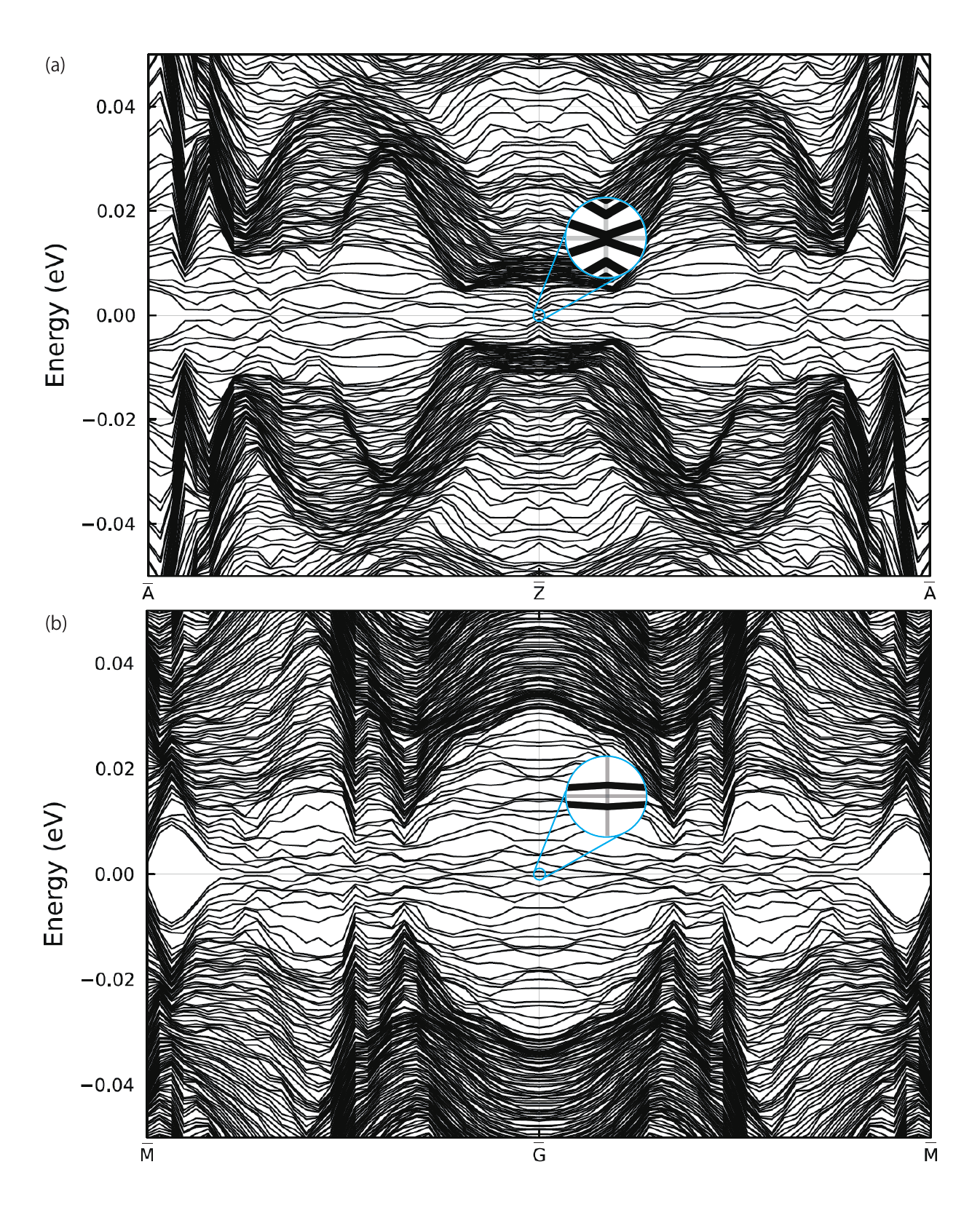}
    \centering
    \caption{$(\bar 110)$ and $(1\bar 10)$ surface states of the $A_u$ superconducting state on (a) $k_z=\pi$ and (b) $k_z=0$. Majorana states appear at $k_z=\pi$, consistent with the nontrivial $\mathbb{Z}_2$ invariant $\nu^{\mathfrak{g}^{\pm}}_{\pi}=1$ in Table II 
    for $U=2.0$ eV, while they do not appear at $k_z=0$ due to $\nu^{\mathfrak{g}^{\pm}}_{0}=0$.}
    \label{fig:slab}
    \end{figure}

    Figure \ref{fig:slab} shows the $(\bar 110)$ and $(1\bar 10)$ surface states of the $A_u$ superconducting state on $k_z=0,\pi$ calculated by diagonalizing the BdG Hamiltonian with the open boundary condition. We set $(a_1, a_2, a_3, a_4) = (0, 0, 0.015, 0.03)$, $H=10$ T for the magnetic field, and $L=64$ for the slab size. We see zero-energy Majorana states at the $\bar Z$ point ($k_z=\pi$), while the states are gapped at the $\bar \Gamma$ point. This is consistent with $\nu^{\mathfrak{g}^{\pm}}_{0}=0$ and $\nu^{\mathfrak{g}^{\pm}}_{\pi}=1$ in Table II for $U=2.0$ eV. Thus, our results on the topological superconductivity and Majorana surface states are verified from the Wannier model. 
    
\end{document}